\documentclass[12pt]{article}
\usepackage{amssymb}
\usepackage{graphicx}
\usepackage{amsmath,amscd}
\def\be{\begin{eqnarray}}
\def\ben{\begin{eqnarray*}}
\def\ee{\end{eqnarray}}
\def\een{\end{eqnarray*}}

\newcommand{\nn}{\nonumber}

\def\D{\mathcal{D}}

\def\myfrac#1#2{\hbox{\large ${#1\over #2}$}}

\def\=:{=\hspace{-.7em}\raisebox{1.1ex}{.}\hspace{.1em}\raisebox{-0.2ex}{.} }
\newcommand{\NF}{N_{\rm F}}
\newcommand{\NC}{N_{\rm C}}
\newcommand{\hs}[1]{\hspace{#1 mm}}

\newcommand {\beq}{\begin{eqnarray}}
\newcommand {\eeq}{\end{eqnarray}}

\makeatletter
\@addtoreset{equation}{section}

\renewcommand{\thefootnote}{\fnsymbol{footnote}}
\makeatother

\makeatletter
\newcommand{\thetablename}{Table}
\def\fnum@table{\thetablename\ \thetable}
\makeatother
\begin{document}
\thispagestyle{empty}
\begin{flushright}
TIT/HEP--539\\
{\tt hep-th/0506257} \\
July, 2005 \\
\end{flushright}
\vspace{3mm}

\begin{center}
{\Large \bf 
1/2, 1/4 and 1/8 BPS Equations in \\  
SUSY Yang-Mills-Higgs Systems}
 
%\vspace{5mm}
{\large -- Field Theoretical Brane Configurations --}
\\[10mm]
\vspace{5mm}

\normalsize
{\large %\bf 
Minoru~Eto}\!\!
\footnote{\it  e-mail address: 
meto@th.phys.titech.ac.jp
}, 
  {\large %\bf 
Youichi~Isozumi}\!\!
\footnote{\it  e-mail address: 
isozumi@th.phys.titech.ac.jp
}, 
  {\large %\bf 
Muneto~Nitta}\!\!
\footnote{\it  e-mail address: 
nitta@th.phys.titech.ac.jp
}  
  {\large %\bf 
~and~~ Keisuke~Ohashi}\!\!
\footnote{\it  e-mail address: 
keisuke@th.phys.titech.ac.jp
} 

\vskip 1.5em

{ \it Department of Physics, Tokyo Institute of 
Technology \\
Tokyo 152-8551, JAPAN  
 }
\vspace{12mm}

%{\bf Abstract}\\[5mm]
%{\parbox{13cm}{\hspace{5mm}
%%%%%%%%%%%%%%%%%%%%%%%%%%%%%%%%%%%%%%%%%%%%%%%%%%
\abstract{
We systematically classify 
1/2, 1/4 and 1/8 BPS equations 
in SUSY gauge theories in $d=6, 5, 4, 3$ and $2$ 
with eight supercharges, 
with gauge groups and matter contents being arbitrary. 
Instantons (strings) and vortices (3-branes) 
are only allowed 1/2 BPS solitons in 
$d=6$ with ${\cal N}=1$ SUSY.  
We find two 1/4 BPS equations and 
the unique 1/8 BPS equation in $d=6$
by considering configurations made of 
these field theory branes. 
All known BPS equations are rederived while 
many new 1/4 and 1/8 BPS equations are found 
in dimension less than six 
by dimensional reductions. 
}
%%%%%%%%%%%%%%%%%%%%%%%%%%%%%%%%%%%%%%%%%%%%%%%%%%
%}}
\end{center}
\vfill
\newpage
\setcounter{page}{1}
\setcounter{footnote}{0}
\renewcommand{\thefootnote}{\arabic{footnote}}

%%%%%%%%%%%%%%%%%%%%%%%%%%%%%%%%%%%%%%%%%%%%%%%%%%%%%%%%%%%%%%%%%%%%%%%%%%%%

%%%%%%%%%%%% text start %%%%%%%%%%%%%%%%%%%

%%%%%%%%%%%%%%%%%%%%%%%%%%%%%%%%%%%%%%%%%%%%%%%%%%%%%%%%%%%%%%%%%%%%%%%%%%%%%
\tableofcontents
\addtocontents{toc}{\vspace{-4mm}}
\section{Introduction}
\addtocontents{toc}{\vspace{-4mm}}
Bogomol'nyi-Prasad-Sommerfield (BPS) states \cite{Bogomolny:1975de} 
are the most important ingredients for recent developments 
in non-perturbative aspects of 
supersymmetric (SUSY) field theory, string theory and M-theory. 
This is because they saturate lower energy bound called 
the Bogomol'nyi bound, 
break/preserve half supercharges, 
belong to short multiplets of supersymmetry, 
and therefore are non-perturbatively stable \cite{Witten:1978mh}. 
Their masses (tension) appear as central (tensorial) charges 
in the SUSY algebras. 
D-branes are such objects in string theory~\cite{Polchinski:1998rq}. 
D-brane configurations made of various dimensional 
D-branes and NS5-branes are very powerful tools to study 
non-perturbative aspects of SUSY field theory realized 
on some D-branes~\cite{brane-config}. 
These brane configurations can be obtained 
from configuration made of M5- and M2-branes in M-theory 
through various dualities.
Therefore classifying possible configuration of 
M5- and M2-branes is important. 
Such classification including possible angles between branes 
was established in \cite{Ohta:1997fr}.

In the field theory side, there exist four kinds of 
fundamental 1/2 BPS solitons \cite{MantonSutcliffe,Rajaraman:1982is};
instantons~\cite{Belavin:1975fg,Atiyah:1978ri}\footnote{
Instantons are 1/2 BPS states if 
embedded into four-dimensional Euclidean space in 
$d=5$, ${\cal N}=2$ (${\cal N}=1$)
SUSY gauge theories with sixteen (eight) supercharges.
}, 
monopoles~\cite{Bogomolny:1975de,Goddard:1976qe,Nahm,Auzzi:2004if}, 
vortices~\cite{ANO}--\cite{Eto:2006uw} and 
domain walls~\cite{Cvetic:1991vp}--\cite{Hanany:2005bq},  
which are of co-dimension four, three, two and one, respectively.\footnote{
In addition to these fundamental solitons, 
there exist several associated solitons: 
they are an instanton and a monopole with electric flux,  
called a dyonic instanton \cite{Lambert:1999ua}  
and a dyon \cite{Julia:1975ff}, respectively, 
and stationary time-dependent lumps and kinks,  
called Q-lumps~\cite{Q-lumps} and 
Q-kinks \cite{Abraham:1992vb}, respectively. 
Here lumps~\cite{lumps} (equivalently
sigma-model instantons~\cite{sigmamodel-instantons} 
in Euclidean two dimensional space) 
can be obtained from semilocal vortices \cite{semilocal} 
in the strong gauge coupling limit. 
}
These BPS solitons can be regarded as 
``field theory branes". 
In fact, recently, many kinds of 
1/4 BPS composite states of several BPS solitons, 
resembling with D-brane configurations in string theory, 
have been found in 
SUSY gauge field theories (or hyper-K\"ahler sigma models) with eight supercharges 
in $d=3,4,5$~\cite{Gauntlett:2000de}--\cite{Portugues:2002ih}.
Further study of similarity between 
branes in string/M theory and branes in field theory 
is earnestly desired. 
However it has not been classified yet 
what kinds of composite solitons 
are allowed for BPS states 
in field theories with eight supercharges.

The purpose of this paper is 
systematically classifying 
possible 1/2, 1/4 and 1/8 BPS equations 
in SUSY gauge theories in $d=6, 5, 4, 3$ and $2$ with eight supercharges, 
with gauge groups and matter contents being arbitrary. 
First of all we find that only 
instantons (strings) and vortices (3-branes) are allowed 
for 1/2 BPS states in $d=6$,  
by classifying projection operators on fermions.  
Second, intersection rules for 
orthogonal instanton-strings and vortex-3-branes 
are obtained. 
By considering orthogonally intersecting field theory branes,  
we find that all possible 1/4 and 1/8 BPS field theoretical 
brane configurations 
in $d=6$ are summarized in Table \ref{summary_table}. 
\begin{table}
\begin{center}
\caption{\small 
1/4 BPS and 1/8 BPS configurations in $d=6$.
Here ``$\bigcirc$" denotes the world-volume directions 
of the solitons, whereas ``$\times$" denotes their codimensional directions.
Throughout this paper, 
we use the following Roman uppercases to represent various solitons: 
I: instanton,  M: monopole, 
H: Hitchin vortex, N: Nahm wall,
V: vortex, W: domain wall.
The indices on these letters imply that those solitons 
extend to various directions.
}
\label{summary_table}
%%%%%%%%%%%%%%%%
\begin{eqnarray*}
\begin{array}{c|cccccc}
 1/4 \;\; {\rm IVV}&0 &1 &2 &3 &4 & 5 \\
\hline
{\rm Instanton}&\bigcirc &\times &\times &\times &\times &\bigcirc  \\
{\rm Vortex}&\bigcirc &\times &\times &\bigcirc &\bigcirc &\bigcirc \\
{\rm Vortex}&\bigcirc &\bigcirc &\bigcirc &\times &\times &\bigcirc  
\end{array}
\hs{5}
\begin{array}{c|cccccc}
1/4 \;\; {\rm VVV} &0 &1 &2 &3 &4 &5 \\
\hline
{\rm Vortex}&\bigcirc &\bigcirc &\times &\times &\bigcirc &\bigcirc \\
{\rm Vortex}&\bigcirc &\times &\bigcirc &\times &\bigcirc &\bigcirc \\
{\rm Vortex}&\bigcirc &\times &\times &\bigcirc &\bigcirc &\bigcirc
\end{array}
\end{eqnarray*}
\begin{eqnarray*}
\begin{array}{c|cccccc}
1/8 \;\; {\rm IV}^6 &0 &1 &2 &3 &4 &5 \\ \hline
{\rm Instanton}&\bigcirc &\times &\times &\times &\times &\bigcirc  \\
{\rm Vortex}&\bigcirc &\bigcirc &\times &\times &\bigcirc &\bigcirc \\
{\rm Vortex}&\bigcirc &\times &\bigcirc &\times &\bigcirc &\bigcirc \\
{\rm Vortex}&\bigcirc &\times &\times &\bigcirc &\bigcirc &\bigcirc \\
{\rm Vortex}&\bigcirc &\times &\bigcirc &\bigcirc &\times &\bigcirc \\
{\rm Vortex}&\bigcirc &\bigcirc &\times &\bigcirc &\times &\bigcirc \\
{\rm Vortex}&\bigcirc &\bigcirc &\bigcirc &\times &\times &\bigcirc
\end{array}
\end{eqnarray*}  
\end{center}
\end{table}
%%%%%%%%%%%%%%%%%%
Third, we derive two 1/4 BPS equations and 
the unique 1/8 BPS equation in $d=6$ corresponding to 
these brane configurations.\footnote{
We first assume minimal kinetic term for hypermultiplets. 
We can consider hyper-K\"ahler nonlinear sigma models  
whose target spaces are  
hyper-K\"ahler quotients \cite{HKquotient} of the flat spaces, 
by taking strong gauge coupling limit.  
BPS equations for these associated 
hyper-K\"ahler nonlinear sigma models 
are also available. 
}  
Fourth, all BPS equations in dimensions less than six 
are derived from them by 
the Scherk-Schwarz and/or ordinary dimensional reductions.\footnote{
$d=6$, ${\cal N}=1$ SUSY reduces to  
$d=5$, ${\cal N}=1$; 
$d=4$, ${\cal N}=2$; 
$d=3$, ${\cal N}=4$ 
or $d=2$, ${\cal N}=(4,4)$ SUSY. 
}  
All the previously known BPS equations are rederived for particular 
choice of gauge group and representations of matter hypermultiplets, 
while several new 1/4 and 1/8 BPS equations are found. 
Some of these equations describe 
monopoles and/or domain walls also. 
Therefore all the field theory branes are 
unified in $d=6$ 
into instanton-strings and vortex-3-branes, 
as if D-branes, fundamental string and NS5-branes are unified into 
M2- and M5-branes in eleven dimensional M-theory. 

We also discuss electrically charged solitons. 
They are new solitons traveling 
with the speed of light in one direction in $d=6$, which we call wavy solitons.
When dimensionally reduced along that direction, 
they become stationary solitons in dimensions less than six. 
They are dyonic solitons including dyonic instantons \cite{Lambert:1999ua} 
(dyons \cite{Julia:1975ff}) in $d=5$ ($d=4$).

When 1/2 BPS instantons, monopoles and Hitchin-type vortices 
(in the Hitchin system) 
are constructed in systems coupled with the Higgs fields,
they become 1/4 BPS composite states of 
instantons accompanied with vortices,
those of monopoles with vortices and domain walls,  
and those of Hitchin vortices with domain walls, 
respectively. 
In Abelian gauge theory, 
they carry negative instanton, monopole and Hitchin charges 
and these charges contribute negatively 
to the total energy at intersections of solitons. 
The first one was found in \cite{Eto:2004rz} 
and was called an ``intersecton". 
The second one was found in \cite{INOS3} and 
later was called a ``boojum" \cite{Sakai:2005sp}.
The last one is a domain wall junction 
with junction charge 
(interpreted as the Hitchin charge)~\cite{Eto:2005cp}. 
In this paper we point out that these 1/4 BPS states are 
related by dimensional reductions, 
similarly to the well-known descent relation between 
instantons, monopoles and Hitchin vortices.

In this paper 
we discuss only 
projections orthogonal to each other, implying 
orthogonal branes (in the absence of fluxes).\footnote{
In the presence of magnetic fluxes they can have angle 
even though projections are orthogonal \cite{INOS3,Eto:2005cp}.
We do not exclude this possibility of angles between branes 
in this paper. 
} 
We do not discuss angles between branes 
as was done in \cite{Ohta:1997fr} 
for M2- and M5-branes in M-theory.
We leave it as a future problem.
We believe that this paper provides 
a guide for pursuing 
similarity between branes in string/M-theory 
and field theory branes.

This paper is organized as follows. 
In Sect.~2, we present Lagrangian in $d=6$ and 
sets of projection operators for 
1/2, 1/4 and 1/8 BPS states.  
Sect.~3, 4 and 5 deal with 1/2, 1/4 and 1/8 BPS states,
respectively. 
Sect.~6 is devoted to conclusion and discussion. 

Before closing the introduction we summarize  
previously known composite BPS solitons in 
SUSY gauge theories and hyper-K\"ahler 
nonlinear sigma models with eight supercharges: 

\begin{enumerate}
\item
[[${1\over 4}$ WV]] Lump-strings ending on a wall was found 
in a hyper-K\"ahler sigma model~\cite{Gauntlett:2000de}. 
It was promoted to a vortex-string ending on a wall 
in $U(1)$ gauge theory~\cite{Shifman:2002jm}. 
These composite states are 1/4 BPS.

\item
[[${1\over 4}$ VM]] A monopole in the Higgs phase is attached by vortices.
The total system becomes a composite 1/4 BPS state of 
a monopole and vortices~\cite{vm}. 

\item
[[${1\over 4}$ WVM]] 
Even if the above [${1 \over 4}$ WV] 
and [${1 \over 4}$ VM] are simultaneously considered, 
a composite state of walls, vortices and monopoles 
is still 1/4 BPS~\cite{INOS3}. 
Moduli were completely determined and all the exact solutions were  
obtained in the strong gauge coupling limit 
(reducing to hyper-K\"ahler sigma model). 

\item
[[${1\over 4}$ IVV]] 
Instanton-particles in $d=5$ are stabilized in the Higgs phase, 
if they are supported by a vortex sheet~\cite{Eto:2004rz}. 
The 1/4 BPS equation for this also contains 
intersecting vortices with intersecting point carrying 
an instanton charge.
Dimensional reductions of this equation to $d=4,3,2$  
reduce to 1/2, 1/4 BPS equations known so far, 
except for [${1 \over 4}$ VVV] below.

\item
[[${1\over 4}$ HWW]] 
Walls were found to constitute a junction 
as 1/4 BPS states in $d=4$, ${\cal N}=1$ SUSY 
theories~\cite{Gibbons:1999np}.
A wall junction carries a junction charge 
in addition to two wall charges.
Some exact solutions are known~\cite{Oda:1999az}. 
As first example in theories with eight supercharges, 
intersecting walls were found 
in a hyper-K\"ahler nonlinear sigma model~\cite{Gauntlett:2000bd}.
A ${\bf Z}_3$ symmetric wall junction 
was constructed 
in $d=4$, ${\cal N}=2$ $U(1)$ gauge theory~\cite{Kakimoto:2003zu}. 
All solutions of wall junctions, called wall webs, have 
been recently found~\cite{Eto:2005cp,Eto:2005fm}  
where all the exact solutions have been obtained 
in the strong gauge coupling limit. 

\item
[[${1\over 4}$ VVV]]
A set of 1/4 BPS equations for triply intersecting lump-strings 
was found in $d=4$, ${\cal N}=2$ hyper-K\"ahler 
nonlinear sigma models~\cite{Naganuma:2001pu}. 
It was lifted in \cite{Portugues:2002ih} 
up to intersecting lump-membranes in $d=5$, ${\cal N}=1$ theory. 
Promoting them to gauge theories have not been done yet. 
Only this set of equations cannot be obtained from the above 
[${1 \over 4}$ IVV] 
by dimensional reduction.

\item
[[${1\over 8}$ Q-VVV]] 
Only one set of 1/8 BPS equations has been known so far 
in theories with eight supercharges. 
It is time dependent extension of the above [${1 \over 4}$ VVV], 
describing triply intersecting Q-lump-strings 
(membranes)~\cite{Naganuma:2001pu,Portugues:2002ih}.

\end{enumerate}

{\bf Note Added. }
When we were preparing the manuscript of this paper 
we were informed from 
Kimyeong Lee and Ho-Ung Yee
that there exist some overlap 
between their paper~\cite{LeeYee} 
and this paper. 
They also derived 1/8 BPS equations. 
They constructed a perturbative solution 
for one of 1/8 BPS equations 
and discussed 1/4 BPS dyonic solitons 
in $d=4$.

\bigskip

%%%%%%%%%%%%%%%%%%%%%%%%%%%%%%%%%%%%%%%%%%%%%%%%%%%%%%%%%%%%%%%%%%%%%%%%%%%%%
\section{Models and SUSY Projection Operators}  \label{sect.2}
%%%%%%%%%%%%%%%%%%%%%%%%%%%%%%%%%%%%%%%%%%%%%%%%%%%%%%%%%%%%%%%%%%%%%%%%%%%%%
\addtocontents{toc}{\vspace{-2.3mm}}
\subsection{Models}
\addtocontents{toc}{\vspace{-2.3mm}}

We begin with the 6 dimensional supersymmetric gauge theory with 
eight supercharges (${\cal N}=1$ in 6 dimensions).
The SUSY multiplets which we will consider in this paper are vector multiplets
and hypermultiplets. The physical fields contained in the vector multiplet are
the gauge field $W_M$ ($M=0,1,2,3,4,5$) and the gaugino $\lambda^i$. 
The gaugino $\lambda^i$ belongs to a doublet of the $SU(2)_R$ ($i=1,2$)
and is the $SU(2)$-Majorana Weyl spinor, namely $\gamma_7\lambda^i=\lambda^i$
and $\lambda^i=C\varepsilon^{ij}(\bar\lambda_j)^T$. 
Here $\gamma_7$ is defined by $\gamma_7 = \gamma^{012345}$ and 
$C$ is the charge conjugation matrix in 6 dimensions\footnote{
Our convention for the metric is 
$\eta_{MN} = {\rm diag}(+,-,\cdots,-)$.
The gamma matrices satisfy 
the Clifford algebra $\{\gamma_M,\gamma_N\} = \eta_{MN}$
and the charge conjugation matrix is 
defined by $C^{-1}\gamma_MC = \gamma_M^T$ 
and $C^T = -C$. 
}.
The physical fields in the hypermultiplets are 
scalar fields $H^{iA}$ which are $SU(2)_R$ doublets and the 
hyperinos $\psi^A$ (Dirac spinor) whose chirality is $\gamma_7\psi^A=-\psi^A$.
The indices $A,B,\cdots$ 
stand for both the gauge group and the flavor symmetry.
Then the bosonic Lagrangian with the minimal kinetic terms in 6 dimensions 
is given by
\begin{eqnarray}
{\cal L}_6 =-{1\over 4g_I^2}F^I_{MN}F^{IMN }+\left({\cal D}_M H^{iA}\right)^*
{\cal D}^M H^{iA}-{1\over 2g_I^2}(Y^{I}_a)^2,
\label{lag6}
\end{eqnarray}
where we define
\begin{eqnarray}
Y^{I}_a =g^2_I\left[\zeta ^I_a-(H^{iA})^*(\sigma _a)^i{}_j(T_I)^A{}_BH^{jB}\right].
\end{eqnarray}
We imply the summation over repeated indices. 
Here $I$ is the index for generators $T_I$ of the gauge group, 
and $g_I$ denotes the gauge coupling constant 
taking the same value for each group 
in the product gauge group. 
In our conventions, a covariant derivative and a field strength 
are given by ${\cal D}_M=\partial _M+iW_M$ and 
$F_{MN}=-i[{\cal D}_M,\,{\cal D}_N]$ (with $X=X^IT_I$), respectively.
Real constants $\zeta^I_a$ ($a=1,2,3$), called 
the Fayet-Iliopoulos (FI) parameters, 
can have nonzero values only for $I$ corresponding 
to the Abelian subgroups. 
Masses for hypermultiplets are prohibited in $d=6$. 
(They can have masses in lower dimensions as shown below.) 

The SUSY transformations for the spinor fields 
in 6 dimensions are given by
\begin{eqnarray}
 \delta _\varepsilon \lambda ^i=\myfrac12\gamma ^{\mu \nu }F_{\mu \nu }
 \varepsilon ^i+Y_a(i\sigma _a)^i{}_j\varepsilon ^j,\quad
\delta _\varepsilon \psi =
-\sqrt{2}i\gamma ^\mu {\cal D}_\mu H^i\epsilon _{ij}\varepsilon ^j .
\label{susy_transf}
\end{eqnarray}
Here, a parameter $\varepsilon ^i$ for the SUSY transformation is 
an $SU(2)$ Majorana-Weyl spinor which satisfies 
$\varepsilon ^i=C\epsilon ^{ij}(\bar \varepsilon _j)^{\rm T},\ 
\gamma _7\varepsilon ^i=+\varepsilon ^i$.

We can obtain the $d (< 6)$ dimensional 
supersymmetric gauge theory with eight supercharges, 
by performing the Scherk-Schwarz (SS) and/or 
the trivial dimensional reductions
($6-d$)-times from the 6 dimensional theory, 
after compactifying the $p$-th ($p=5,4,\cdots,d$) 
direction to $S^1$ with radius $R_p$. 
We consider the twisted boundary condition for the
dimensional reduction along the $x^p$-direction,  
given by 
\begin{eqnarray}
 H^{iA}(x^\mu,x^p+2\pi R_p) = 
 H^{iA}(x^\mu, x^p)e^{i\alpha _{pA}},\quad \left(|\alpha _{pA}|\ll 2\pi\right),
  \label{SS-red.}
\end{eqnarray}
where $\mu (=0,\cdots,d-1)$ is a spacetime index in $d$ dimensions.
If we consider effective action at sufficiently low energy, 
we can ignore the infinite towers of the Kaluza-Klein modes and 
 have the lightest fields as functions 
 of the $d$ dimensional spacetime coordinates 
\begin{eqnarray}
&&W_\mu(x^\mu,x^p) \rightarrow W_\mu(x^\mu),\quad 
W_p(x^\mu,x^p) \rightarrow -\Sigma_p (x^\mu),\\
&&H^{iA}(x^\mu,x^p)\rightarrow {1\over \prod_p\sqrt{2\pi R_p}}
H^{iA}(x^\mu)\exp\left({i\sum_pm_{pA}x^p}\right),\quad 
m_{pA}\equiv {\alpha _{pA}\over 2\pi R_p}.
\end{eqnarray}
Integrating the 6 dimensional Lagrangian in Eq.(\ref{lag6})
over the $x^p$-coordinates, 
we obtain the $d$ dimensional Lagrangian
\be
{\cal L}_d &=& 
- {1\over 4g_I^2}F^I_{\mu\nu}F^{I\mu\nu} + 
\frac{1}{2g_I^2}\D_\mu\Sigma_p^I\D^\mu\Sigma_p^I + 
\left({\cal D}_\mu H^{iA}\right)^*{\cal D}^\mu H^{iA} \nonumber\\
&& +{1\over 4g_I^2}([\Sigma_p, \Sigma_q]^I)^2
- {1\over 2g_I^2}(Y^{I}_a)^2 - 
\left|H^{iA}m_{pA} - \Sigma^I_p(T_I)^A{_B}H^{iB}\right|^2,
\ee
where we have redefined 
the gauge couplings and the FI parameters in $d$ dimensions
from 6 dimensions as
$g_I^2\rightarrow \left(\prod_p2\pi R_p\right)g_I^2$,
$\zeta ^I_a\rightarrow \zeta ^I_a/ \left(\prod_p2\pi R_p\right)$.
By construction one finds that 
there exist $(6-d)$ real mass parameters for hypermultiplets 
in $d$ dimensions.

In this paper we are mainly interested in composite states of solitons
living inside the Higgs branches of gauge theories with eight supercharges,
in particular when we consider vortices and/or domain walls.  
Therefore we prepare sufficiently large number of hypermultiplets. 
For the massless hypermultiplets, 
the Higgs branches are hyper-K\"ahler quotient~\cite{HKquotient}  
given by ${\cal M}_{\rm vac.} = \{H^{iA}\ |\ Y^I_a=0\}/G$ 
where $G$ denotes the gauge group. 
The real dimension of the Higgs branch
is $\dim {\cal M}_{\rm vac.} = 4(n_H - \dim G)$ with 
$n_H$ the number of hypermultiplets.
When we introduce masses in lower dimensions 
by the SS dimensional reductions, 
the Higgs branch of vacua is lifted by masses 
except for fixed points of the $U(1)$ Killing vectors~\cite{Alvarez-Gaume:1983ab} 
induced by the $U(1)$ actions in Eq.~(\ref{SS-red.}). 
In the strong gauge coupling limit, 
the model reduces to a hyper-K\"ahler nonlinear sigma model 
on the hyper-K\"ahler manifold ${\cal M}_{\rm vac.}$ 
with a potential given by the square of 
the Killing vectors~\cite{Alvarez-Gaume:1983ab}.

A typical model often considered so far  
is $U(\NC)$ gauge theory coupled to $\NF$ ($\geq \NC$) hypermultiplets in 
the fundamental representation 
with one triplet of FI parameters.
There exists the unique Higgs vacuum for $\NF = \NC$, 
called the color-flavor locking phase, 
up to gauge transformation. 
Massless vacua for $\NF > \NC$ are given by the cotangent bundle over 
the complex Grassmann manifold 
$G_{\NF,\NC} \simeq SU(\NF)/[SU(\NC) \times SU(\NF-\NC) 
\times U(1)]$~\cite{HKquotient}. 
Introducing non-degenerate masses, 
only $\NF!/[\NC! \times (\NF-\NC)!]$ discrete 
points remain as vacua~\cite{ANS}. 
In the strong gauge coupling limit, 
the model reduces 
to the massive hyper-K\"ahler nonlinear sigma model on 
$T^* G_{\NF,\NC}$.  
Another model often considered is 
a hyper-K\"ahler nonlinear 
sigma model on the ALE space.
It can be obtained in the strong coupling limit from 
$U(1)^N$ gauge theory 
coupled to $(N+1)$ hypermultiplets with suitable $U(1)$ charges 
and $N$ triplets of FI-parameters~\cite{Gibbons:1996nt}. 
The $N$ centers in the ALE space are vacua for massive case. 
In \cite{Eto:2005wf} we have discussed domain walls 
in the hyper-K\"ahler nonlinear sigma model on 
the cotangent bundle over the Hirzebruch surface $F_n$, 
which can be obtained 
by $U(1)^2$ gauge theory coupled to four hypermultiplets 
with suitable $U(1)^2$ charges and 
with two parallel triplets of FI-parameters.

%%%%%%%%%%%%%%%%%%%%%%%%%%%%%%%%%%%%%%%%
\subsection{Projection operators}
\addtocontents{toc}{\vspace{-4mm}}

There exist several BPS solitons 
in the supersymmetric gauge theories.
The typical property of BPS solitons is that they partially preserve
supersymmetry.
The most elementary objects are 1/2 BPS solitons preserving
a half of eight supercharges. 
They are characterized by supercharges which
solitons preserve. 
An $SU(2)$ Majorana-Weyl spinor parameter $\varepsilon$ for unbroken SUSY 
is specified by the equation 
\begin{eqnarray}
 (\Gamma\varepsilon )^i=\pm\varepsilon ^i,\quad {\rm Tr}(\Gamma)=0,
\label{projection}
\end{eqnarray}
where $\Gamma$ is an operator satisfying $\Gamma^2={\bf 1}_8\otimes {\bf 1}_2$.
The traceless condition is required for choosing a half of supercharges by
operator $\Gamma$.
Since the SUSY transformation parameter $\varepsilon^i$ is 
the $SU(2)$ Majorana-Weyl spinor, 
$\Gamma$ has to satisfy the following two conditions
for consistency,  
\begin{eqnarray}
 B^{-1}\Gamma B= \Gamma ^*,\quad [\gamma _7,\,\Gamma]=0,
\label{consistency}
\end{eqnarray}
with $B^{ij} \equiv \gamma^0C\epsilon^{ij}$.
If a soliton with codimension $D$ exists in 6 dimensions,
the Lorentz symmetry $SO(1,5)$ is broken to $SO(1,5-D)\otimes SO(D)$.
Therefore the operator $\Gamma$ has to be invariant under 
$SO(1,5-D)\otimes SO(D)$ transformation.
From the second condition in Eq.(\ref{consistency}) 
the operator $\Gamma$ has to consist of 
a product of even number
of the $\gamma^M$ matrices.
Therefore, the codimension $D$ of the 1/2 BPS solitons
must be even in 6 dimensions. Combining this fact with 
the first condition in Eq.(\ref{consistency}),
we find that possible operators are given by
\begin{eqnarray}
 && \text{type I\!Va}:\quad 
  \Gamma = \gamma ^{nm}i\sigma _a,\quad \label{4a} \\
 && \text{type I\!Vb}:\quad
  \Gamma = \gamma ^{0n},  \label{4b}
\end{eqnarray}
where $n,m=1,2,\cdots,5$ and $a=1,2,3$.
As we will see in the following section,
the first operator which we call type I\!Va implies 
existence of 1/2 BPS vortices of codimension two 
in the $x^n$-$x^m$ plane,   
whose world-volume extends to the time direction and 
the four space directions except for $x^n$ and $x^m$.  
Vortices are 3-branes in $d=6$. 
On the other hand
the second operator which we call type I\!Vb 
implies existence of 1/2 BPS instantons of codimension four, 
whose world-volume extends to the time direction and 
one space direction $x^n$.
Instantons are strings (1-branes) in $d=6$.  
Thus we conclude that the elementary 1/2 BPS solitons in 6 dimensions are
vortices and instantons, 
which are  3-branes and strings, respectively.

When there exist two or more 1/2 BPS solitons 
preserving 1/2 supercharges different from each other,
the composite state of these solitons breaks more than four 
supercharges out of eight.
If we 
impose different 1/2 BPS projection operators for the Killing spinors
with general angles, the SUSY will be completely broken. 
However, composite solitons 
determined by orthogonal 1/2 BPS projection operators 
can preserve 1/4 or 1/8 SUSY, 
like orthogonal D-brane/M-brane configurations in superstring/M theory.
In this article 
we concentrate on composite states 
of the orthogonal 1/2 BPS solitons.
With respect to the projection operator $\Gamma$ for the supertransformation
parameter in Eq.(\ref{projection}), 
there exist two sets of $\Gamma$'s for 1/4 BPS states. 

One type which we call type I\!Ia is a set of three projections 
\be
\text{type I\!Ia}:
\quad \Gamma =
\left\{
\gamma^{12}i\sigma_3,\ 
\gamma^{23}i\sigma_1,\ 
\gamma^{31}i\sigma_2
\right\}
\label{type A}
\ee
implying three orthogonal vortex-3-branes. 
The projections in (\ref{type A}) correspond 
to vortex-3-branes in the $x^1$-$x^2$ plane 
whose world-volume extends to 
the $x^0$, $x^3$, $x^4$ and $x^5$-directions, 
vortex-3-branes in the $x^2$-$x^3$ plane extending to 
the $x^0$, $x^1$, $x^4$ and  $x^5$-directions, 
and vortex-3-branes in the $x^3$-$x^1$ plane 
extending to the $x^0$, $x^2$, $x^4$ and $x^5$-directions, 
respectively (see the second table in p.3). 
Notice that all projections in (\ref{type A}) 
commute each other 
(for example $\left[\gamma^{12}i\sigma_3,\gamma^{23}i\sigma_1\right]=0$) 
and that 
the product of all of them is proportional to identity 
($\gamma^{23}i\sigma_1\cdot\gamma^{31}i\sigma_2\cdot\gamma^{12}i\sigma_3 = -{\bf 1}$).
These facts imply that the set in Eq.(\ref{type A}) 
project out 3/4 of supercharges,
so that composite states of the orthogonal three vortices are 1/4 BPS states
in 6 dimensions. 

The other set of operators $\Gamma$'s 
which we call type I\!Ib is given by
\be
\text{type I\!Ib}:\quad
\Gamma=
\left\{
\gamma^{05},\ 
\gamma^{12}i\sigma_3,\ 
\gamma^{34}i\sigma_3
\right\}.
\label{type B}
\ee
This set of operators leads to 1/4 BPS composite states 
of 1/2 BPS instanton-strings 
in the $x^1$-$x^2$-$x^3$-$x^4$ space 
extending to the $x^0$ and $x^5$-directions,  
vortex-3-branes in the $x^1$-$x^2$ plane extending to 
$x^0$, $x^3$, $x^4$, $x^5$-directions 
and vortex-3-branes in the $x^3$-$x^4$ plane 
extending to the $x^0$, $x^1$, $x^2$, $x^5$-directions 
(see the first table in p.3).

Finally, we obtain the unique set of operators $\Gamma$'s 
for 1/8 BPS composite states, 
which we call the type I:  
\be
\text{type I}:\quad
 \Gamma = 
\left\{
\gamma^{05},\ 
\gamma^{12}i\sigma_3,\ 
\gamma^{34}i\sigma_3,\ 
\gamma^{23}i\sigma_1,\ 
\gamma^{14}i\sigma_1,\ 
\gamma^{31}i\sigma_2,\ 
\gamma^{24}i\sigma_2
\right\}. 
 \label{type-I}
\ee
These imply instanton-strings and vortex-3-branes 
in various directions (see the third table in p.3). 
We can easily show that all of these commute each other
and that only three of them are independent operators; 
1/8 BPS is realized by taking 
arbitrary set of three projections 
except for the sets in type I\!Ia/I\!Ib, 
for example 
$\gamma^{05},\gamma^{12}i\sigma_3$ and $\gamma^{23}i\sigma_1$.

To close this section, it is important to study properties 
of surviving supercharges under the above projections. 
It determines SUSY in the effective theory on 
the world-volume of solitons, 
and which geometry (Riemann, K\"ahler or hyper-K\"ahler) 
is realized as the moduli space of solitons.
In two-dimensional space-time $x^0$-$x^5$, chirality of supercharges 
can be determined by $\gamma^{05}$ which is in a reducible representation.
Thus it is convenient to classify the above projections by eigenvalues
of $\gamma^{05}$ on the projected spaces,
even in cases that world-volume of solitons is not two dimensional.        
In the cases of type I\!Vb and type I\!Ib, these sets of 
projection operators  contain just $\gamma^{05}$, and thus, supercharges
projected by these operators are all chiral (or anti-chiral).
On the other hand, in the case of type  I\!Va and type I\!Ia,
numbers of surviving chiral and anti-chiral supercharges 
under these projections are the same. 
This can be proved by 
taking a trace of $\gamma ^{05}$ on a projected space of the Killing spinor, 
for instance in type I\!Va, 
\begin{eqnarray}
 {\rm Tr}\left(\gamma ^{05}P\right)=0, \quad {\rm  with~}
P={{\bf 1} \pm \gamma ^{12}i\sigma _3\over 2}{{\bf 1}+\gamma _7\over 2}. 
\end{eqnarray}
Therefore, if we consider solitons with two-dimensional
world volume $x^0$-$x^5$,
a two-dimensional effective action on 
such solitons has the following supercharges 
\begin{eqnarray}
 && {\rm type~I\!Va:~} {\cal N}=(2,2),\quad {\rm type~I\!Vb:~} {\cal N}=(4,0),\nn\\
 && {\rm type~I\!Ia:~} {\cal N}=(1,1),\quad {\rm type~I\!Ib:~} {\cal N}=(2,0),\nn\\
 && {\rm type~I:~} {\cal N}=(1,0),
\end{eqnarray}
respectively.

%%%%%%%%%%%%%%%%%%%%%%%%%%%%%%%%%%%%%%%%%%%%%%%%%%%%%%%%%%%%%%%%%%%%%
\section{1/2 BPS Systems}  \label{sect.3}
%%%%%%%%%%%%%%%%%%%%%%%%%%%%%%%%%%%%%%%%%%%%%%%%%%%%%%%%%%%%%%%%%%%%%
\addtocontents{toc}{\vspace{-2.3mm}}
In this section we will derive diverse 1/2 BPS equations of solitons 
(instantons, monopoles, vortices and domain walls and so on) in 
$d=6,5,4,3,2,1$ dimensions. 
Since there are two kinds of 1/2 BPS projections, 
the type I\!Va (vortices) and the type I\!Vb (instantons) in 6 dimensions 
as we explained in Eqs.~(\ref{4a}) and (\ref{4b}) in the previous section, 
1/2 BPS systems in all dimensions 
will be classified into two series.

In subsections \ref{Sec.IVb} and \ref{Sec.IVa}
we will first begin with the 1/2 BPS equations for instantons and 
vortices in 6 dimensions, respectively, 
and then derive the 1/2 BPS equations for other solitons
via the trivial dimensional reductions and/or
the Scherk-Schwarz dimensional reductions.

%%%%%%%%%%%%%%%%%%%%%%%%%%%%%%%%%%%%%%%%%%%%%%%%%%%%%%%%%%%%%%%%%%%%%
\subsection{Series of the type I\!Vb} \label{Sec.IVb}
%%%%%%%%%%%%%%%%%%%%%%%%%%%%%%%%%%%%%%%%%%%%%%%%%%%%%%%%%%%%%%%%%%%%%
\addtocontents{toc}{\vspace{-2.3mm}}

Let us start with the 1/2 BPS equations 
which are derived by a half of supercharges specified
by the type I\!Vb operator (\ref{4b}) in 6 dimensions
\begin{eqnarray}
\gamma ^{05}\varepsilon ^i=\gamma ^{1234}\varepsilon ^i
= \eta \,\varepsilon ^i, \quad \eta =\pm 1 .
\label{1/2:inst}
\end{eqnarray}
On the space projected by these Killing conditions, 
the representation of gamma matrices is reducible 
and the matrices $\gamma ^5$ and $\gamma ^4$ can be rewritten as 
$\eta \,\gamma ^0$ and $\eta \,\gamma^{123}$, respectively. 
The corresponding 1/2 (anti-) BPS equations 
can be derived by imposing the 1/2 SUSY condition (\ref{1/2:inst})
to the supersymmetric transformation law (\ref{susy_transf}).
The SUSY transformation laws (\ref{susy_transf}) for the gaugino $\lambda^i$ 
and hyperino $\psi^A$ can be rewritten by using 
the Killing conditions (\ref{1/2:inst}) as
\begin{eqnarray}
\!\!\!\!\delta \lambda ^i
\!\!\!\!&=&\!\!\!\! \left[\myfrac12 \sum_{n,m,l=1}^3
\left(F_{nm}-\eta \,\epsilon _{nml}F_{l4} \right) \gamma ^{nm}
+ \sum_{n=1}^4\left(F_{0n}+\eta F_{5n}\right) \gamma ^{0n}
+ \eta F_{05} \right]\varepsilon^i
+Y_a(i\sigma _a)^i{}_j\varepsilon ^j,\\
\!\!\!\!\delta\psi^A \!\!\!\!&=&\!\!\!\!
- \sqrt 2 i \left[\gamma^0 \left(\D_0 +\eta\D_5\right)H^{iA}
+ \gamma^n\D_nH^{iA}\right]\epsilon_{ij}\varepsilon^j,
\end{eqnarray}
respectively. 
By imposing $\delta\lambda^i = 0$ and $\delta\psi^A = 0$, we find 
1/2 BPS equations
\begin{eqnarray}
F_{nm}=\frac \eta 2\,\epsilon _{nmkl}F_{kl},\ Y_a=0,\ {\cal D}_nH^i=0,
\label{SD}\\
 F_{0n}+\eta F_{5n}=0,\ F_{05}=0,\ \left({\cal D}_0+\eta {\cal D}_5\right)H^i=0,
 \label{wave_SD}
\end{eqnarray}
with $n,m,k,l=1,\cdots,4$.
As discussed 
in subsection \ref{sec:wavy},
when we consider solitons with nontrivial configurations of $W_0^I$ or 
time dependent solutions 
(instanton-string with traveling wave),
we need to take into account 
the Gauss law  
\begin{eqnarray}
 {1\over g^2_I}{\cal D}_nF^I_{0n} = 
 2i(T_I)^A{}_BH^{iB}\stackrel{\leftrightarrow }{\cal D}_0(H^{iA})^* 
\end{eqnarray}
in addition to the above BPS
equations.

In the following, 
subsection \ref{sec:inst-and-decen} is devoted to a review 
whereas subsection \ref{sec:wavy} is new.

%%%%%%%%%%%%%%%%%%%%%%%%
\subsubsection{Instantons and their descendants
%dimensional reductions
} \label{sec:inst-and-decen} \addtocontents{toc}{\vspace{-2.3mm}}
%%%%%%%%%%%%%%%%%%%%%%%%

Let us review well-known BPS equations for 
instantons and their descendants obtained 
by dimensional reductions.
Topological charges for these solitons emerge 
in also 1/4, 1/8 BPS states 
as explained in the following sections. 
Especially, these charges exist even 
in Abelian gauge theories and have negative
contributions to the energy of those systems, 
where those negative charges should be 
interpreted as binding energy 
of composite solitons. 

%%%%%%%%%%%%
{\it Instantons}: 
Let us first consider time-independent (static) BPS solutions 
in six dimensions, by ignoring the $x^5$-dependence and 
by setting $W_0=W_5=0$ in Eq.(\ref{wave_SD}),
namely by considering only Eq.(\ref{SD}). 
The third equation in Eq.(\ref{SD}) implies 
$[{\cal D}_n,\,{\cal D}_m]H^{iA}=i(F_{nm})^A{}_BH^{iB}=0$, 
that is, instanton configurations require 
that charged matter should vanish. 
This is consistent with 
the extended Derrick's theorem \cite{MantonSutcliffe} 
which 
insists that there are no finite energy solutions whose codimension is 4 
when the scalar fields have nonzero vacuum expectation value.
Notice that the FI parameters $\zeta_a^I$ should be turned off for
consistency between $Y^a=0$ and $H^{iA}=0$.
Instantons  become essentially the same with those in 
the SUSY gauge theory with sixteen supercharges 
by adding adjoint hypermultiplets implicitly. 
Thus Eqs.~(\ref{SD}) reduce to the ordinary 
(anti) self-dual (SDYM) equations of
1/2 BPS instantons
\be
F_{nm}=\frac \eta 2\,\epsilon _{nmkl}F_{kl}.\label{eq:bps_inst}
\ee
The solutions of the (anti) SDYM equations saturate the BPS energy bound
\begin{eqnarray}
E_{\rm m}&=&\eta \,{\cal I}_{1234}, \nn \\
{\cal I}_{1234}&\equiv &\int d^4x{1\over 4g^2_I}\epsilon _{nmkl}F_{nm}^IF_{kl}^I
={8\pi ^2\over g^2_I}k^I, \label{eq:def-I}
\end{eqnarray}
with $m,n,k,l=1,2,3,4$. 
Here $E_{\rm m}$ is energy for 
static solutions of instanton 
(we ignore the volume of instanton-string along the $x^5$-direction), 
and ${\cal I}_{pqrs}$ denotes
the charge of the instanton localized in the 
$x^p$-$x^q$-$x^r$-$x^s$ space.
As we will see below soon, the instanton charge ${\cal I}_{mnkl}$
reduces to the charges of other solitons when we perform dimensional 
reduction several times. 
We denote the instanton charge after the dimensional reduction
along the $x^k$-direction as ${\cal I}_{mn\check kl}$,  
with the check ``$\check k$" on the index implying 
to omit that index.

%%%%%%%%
\medskip
{\it Monopoles}: 
By performing trivial dimensional reduction along the $x^4$-coordinate,
the (anti) SDYM equations for the 1/2 BPS instantons reduce to 
the 1/2 BPS equations of the monopoles
\begin{eqnarray}
{\myfrac12}\epsilon _{nml}F_{ml}=-\eta \,{\cal D}_n\Sigma _4, \quad (n,m,l=1,2,3)  
\label{eq:monopole}
\end{eqnarray}
This is well-known Bogomol'nyi equation for BPS monopole, 
whose solutions saturate the BPS energy bound 
\begin{eqnarray}
  E_{\rm m}&=&\eta \,{\cal M}_{123}\geq 0,\nn \\
 {\cal M}_{123}& \equiv & {\cal I}_{123\check4} =
-\int d^3x{1\over g^2_I}\epsilon _{nmk}F_{nm}^I{\cal D}_k\Sigma ^I_4
= -\int d^3x{1\over g^2_I}\epsilon _{nmk}\partial _k\left(F_{nm}^I\Sigma ^I_4\right),  \label{eq:def-M}
\end{eqnarray}
with $n,m,k=1,2,3$. 

%%%%%%%%%
\medskip
{\it The Hitchin system}:
By performing trivial dimensional reduction further 
along the $x^2$-direction,
the 1/2 BPS equations (\ref{eq:monopole}) 
of the monopoles reduce to the 1/2 BPS equations
for the Hitchin system, given by 
\begin{eqnarray}
F_{13}=-\eta i[\Sigma _2,\Sigma _4],\quad 
{\cal D}_1\Sigma _2=\eta {\cal D}_3\Sigma _4,\quad 
{\cal D}_1\Sigma _4=-\eta {\cal D}_3\Sigma _2 .\label{eq:bps_hitc}
\end{eqnarray}
Solutions saturate the BPS bound 
\begin{eqnarray}
 E_{\rm m}&=&\eta \,{\cal H}_{13}\geq 0,\nn \\
{\cal H}_{13} &\equiv& {\cal I}_{1\check23\check4} =
\int d^2x{4\over g^2_I}\partial _{[1} 
 \left(({\cal D}_{3]}\Sigma _4^I)\Sigma _2^I\right).
  \label{eq:def-H}
\end{eqnarray}
We call solitons carrying this topological charge with codimension two 
as ``Hitchin vortices". 
It is known that finite energy solutions for the Hitchin system 
exist in compact spaces but not in non-compact spaces, 
because of divergence of energy \cite{Cherkis:2000cj}. 
However finite energy solutions (as domain wall junctions) 
exist in the Higgs phase even in a non-compact space 
\cite{Eto:2005cp,Eto:2005fm}.

%%%%%%%%%
\medskip
{\it The Nahm equation (dual monopole)}: 
The 1/2 BPS equations for the Hitchin system reduce to 
\begin{eqnarray}
{\cal D}_1\Sigma _p= -{\eta \over 2}
\epsilon _{pqr}i[\Sigma _q,\Sigma _r], \quad p,q,r=2,3,4
\label{eq:bps_nahm}
\end{eqnarray}
by trivial dimensional reduction 
along the $x^3$-coordinate.
This is called the (1/2 BPS) Nahm equation. 
The BPS bound is obtained as
\begin{eqnarray}
 E_{\rm m}&=&\eta \,{\cal N}_{1}\geq 0,\nn \\
{\cal N}_{1}& \equiv &{\cal I}_{1\check2\check3\check4} =
 \int dx^1{1\over 3g^2_I}\epsilon _{pqr}\partial _1
 \left(\Sigma _p^I[\Sigma _q,\Sigma _r]^I\right),
 \label{eq:def-N}
\end{eqnarray}
with $p,q,r=1,2,3$.
We call solitons carrying this topological charge with codimension one 
as ``Nahm walls". 
The energy of solutions diverge again in non-compact space, 
but we do not know at this stage if there exist 
finite energy solutions as the Hitchin system in the Higgs phase.

%%%%%%%%%%
\medskip
{\it Equations in zero dimension:} 
If we fully reduce the codimension for instantons, 
we obtain reduced ADHM-like equations 
\begin{eqnarray}
 [\Sigma _p,\,\Sigma _q]={\eta \over 2}\epsilon _{pqrs}[\Sigma _r,\Sigma _s],
 \quad (p,q,r,s=1,\dots,4) .
\end{eqnarray}
One might consider that if this equation would admit a solution 
it described a space-time filling brane in $d=1$ or $2$ 
and gave a model for partial breaking of SUSY~\cite{PSSB} 
in that dimension. 
Unfortunately, 
this equation, however, has no non-trivial solution, 
$[\Sigma _p,\Sigma _q]\not=0$, 
because the Bogomol'nyi bound vanishes in this case. 
This can be confirmed directly by   
\begin{eqnarray}
E_{\rm m}\propto \epsilon _{pqrs}{\rm Tr}\left\{[\Sigma _p,\,\Sigma _q][\Sigma _r,\Sigma _s]\right\}
=-\epsilon _{pqrs}{\rm Tr}\left\{\Sigma _q[\Sigma _p,\,[\Sigma _r,\Sigma _s]]\right\}=0.
\end{eqnarray}
In this paper, we ignore 
the infinite tower of Kaluza-Klein modes as explained before. 
However if we would consider their contributions,
we could obtain the ADHM equations as 
dual of four-dimensional SDYM equations.

%%%%%%%%%%%%%%%%%%%%%%%%%%%%%%%%%%%%%%%%%%%%%%
\subsubsection{Solitons with electric flux}  \label{sec:wavy}
%%%%%%%%%%%%%%%%%%%%%%%%%%%%%%%%%%%%%%%%%%%%%%
\addtocontents{toc}{\vspace{-2.3mm}}

{\it Wavy solitons}: 
Let us next consider the solitons with the electric flux,
by taking Eq.(\ref{wave_SD}) into account in addition 
to Eq.(\ref{SD}). 
Since $H^{iA}$ has to vanish,
Eq.(\ref{wave_SD}) reduces under dimensional reductions, 
taken $(6-d)$-times except for $x^5$, to 
\be
F_{0n} + \eta F_{5n} =0,\quad
\D_0\Sigma_p + \eta\D_5\Sigma_p = 0,\quad
F_{05} = 0,
\label{eq_05}
\ee
with $p=d-1,\cdots,4$ and $n=1,\cdots,d-2$.
For any static instanton 
(monopole, Hitchin vortex or Nahm wall) 
background configurations $W_n^{\rm static}$ and 
$\Sigma _p^{\rm static}$, 
these equations reduce with an appropriate gauge choice to   
\be
 \partial_0 +\eta\partial_5 = 0,\quad
 W_0 + \eta W_5 = 0.
\ee
General solutions for the first equation can be obtained 
by promoting {\it arbitrary} moduli parameters 
$\phi ^i$ of 
solutions for the static BPS equations 
to arbitrary functions of $x^0-\eta x^5$:
\begin{eqnarray}
 \phi ^i\quad \rightarrow \quad \phi ^i(x^0-\eta x^5).
   \label{wavy-sol}
\end{eqnarray} 
This solution indicates that waves on BPS solitons 
propagate with the speed of light along the $x^5$-direction, 
to which those solitons extend,
and oscillation of magnetic objects induce oscillation of
electro-magnetic waves. 
In the case of time-dependent solitons, 
one should solve the Gauss law  
\be
 \D_nF_{0n} +i \sum_{p=d-1}^4\left[\Sigma_p,\D_0\Sigma_p\right]= 0,
  \label{Gauss-wavy}
\ee
in addition to the BPS equations. 
This equation should be solved with respect to $W_0$ ($W_5$). 
Let us consider the case 
that the moduli parameters in (\ref{wavy-sol}) 
are taken to be the center of mass 
(translational zero modes)
$\phi^m$ ($m=1,\cdots,d-1$). 
We thus have 
\begin{eqnarray}
 W_n(x^0,x^5,x^m)&=&W_n^{\rm static}(x^m-\phi ^m(x^0-\eta x^5)),\quad 
 \\
 \Sigma_p (x^0,x^5,x^m)&=& \Sigma_p^{\rm static}(x^m-\phi ^m(x^0-\eta x^5)),\quad 
\end{eqnarray} 
with $n,m=1,\cdots,d-2$ and $p,q = d-1, \cdots, 4$, 
where the second equation is not needed for instantons 
in $d=6$. 
In this case, 
we find that the Gauss law (\ref{Gauss-wavy}) 
is automatically satisfied by setting
\begin{eqnarray}
  W_0 (= -\eta W_5) = -{d\phi ^m\over dx^0}W_m.
\label{eq:sol_wavy}
\end{eqnarray}
The energy of wavy configurations is given by
\begin{eqnarray}
 E=-\eta S_5+\int dx^5E_{\rm m} (\geq 0),
\ee
where 
the fifth component of the Poynting vector $S_5$ defined by
\be
S_5 &=& \frac{1}{g_I^2}\int d^{d-1}x 
 \left( F_{0n}^IF_{5n}^I 
 + \sum_{p=d-1}^4 \D_0\Sigma_p^I\D_5\Sigma_p^I
 \right) \nonumber\\
&=& 
 - \frac{\eta}{2g_I^2}\int d^{d-1}x \left[(F_{0n}^I)^2+(F_{5n}^I)^2
 +\sum_{p=d-1}^4 \left((\D_0\Sigma_p^I)^2 +(\D_5\Sigma_p^I)^2\right)\right] 
\label{poyinting}
\end{eqnarray}
represents the energy excited by waves on solitons.

%%%%%%%%%%%%%
\medskip
{\it Dyonic solitons}: 
When we perform trivial dimensional reduction from equations 
for wavy solitons along the $x^5$-direction, 
the equations for 
the dyonic solitons\footnote{
Electrically charged instanton in $d=5$
is called dyonic instanton \cite{Lambert:1999ua} 
whereas dyon is electrically charged monopole 
in $d=4$ \cite{Julia:1975ff}. 
The dyonic instanton (field theory supertube) in 6 dimensions
with 16 supercharges was obtained by boosting the $x^5$-direction
with the speed of light and reducing that direction \cite{Townsend:2004nc}.} 
in ($d-1$)-dimensions are obtained. 
Eq.(\ref{eq_05}) reduces to 
\begin{eqnarray}
 F_{0n}+\eta {\cal D}_n\Sigma_5=0,\quad 
 \D_0\Sigma_p - i\eta\left[\Sigma_5,\Sigma_p\right] =0,\quad
 {\cal D}_0\Sigma_5 =0.\label{eq:bps_dyon}
\end{eqnarray}
For any instanton background configuration $W_n$, 
these equations 
can be solved by setting 
\be
 \partial_0 = 0,\quad
 W_0 - \eta \Sigma_5 = 0.
 \label{eq_05_5}
\ee
The configuration of the adjoint scalar $\Sigma_5$ is determined by the
Gauss law as
\begin{eqnarray}
 {\cal D}_nF_{0n} + i\sum_{p=d-1}^4 \left[\Sigma_p,\D_0\Sigma_p\right] = 0,
 \quad\rightarrow\quad 
 {\cal D}_n{\cal D}_n\Sigma_5 
 - \sum_{p=d-1}^4\left[\Sigma_p,\left[\Sigma_p,\Sigma_5\right]\right] =0.
\end{eqnarray}
It is known that there exists the unique solution to this equation 
once an instanton configuration 
and an asymptotic value of $\Sigma_5$
are given~\cite{Lambert:1999ua}. 
The Poynting vector given in Eq.(\ref{poyinting}) reduces to the
electric charge of the dyonic instanton as
\begin{eqnarray}
S_5\rightarrow Q_{\rm e}= \frac{1}{g_I^2}
\int d^{d-1}x \left(
F_{0n}^I{\cal D}_n\Sigma ^I_5
+ i\sum_{p=d-1}^4 \D_0\Sigma_p^I\left[\Sigma_p,\Sigma_5\right]^I\right)
= \frac{1}{g_I^2}
\int d^{d-1}x\partial _n\left(F_{0n}^I\Sigma ^I_5\right),
\end{eqnarray}
and the energy density becomes
\be
E=-\eta Q_{\rm e}+E_{\rm m} (\geq 0).
\ee

%%%%%%%%%%%%%%%%%%%%%%%%%%%%%%%%%%%%%%%%%%%%%%%%%%
\subsection{Series of the type I\!Va: vortices and domain walls} \label{Sec.IVa}
%%%%%%%%%%%%%%%%%%%%%%%%%%%%%%%%%%%%%%%%%%%%%%%%%%
\addtocontents{toc}{\vspace{-2.3mm}}
This subsection is devoted to a review. 
Let us next investigate 1/2 BPS solitons which preserve
four supercharges specified by the type I\!Va projection operator
$\gamma^{nm}i\sigma_a$ on the Killing spinors, given in 
Eq.~(\ref{4a}). 
Especially, we deal with the projection
\begin{eqnarray}
 \gamma ^{12}\left(i\sigma _3\varepsilon\right) ^i=\xi \varepsilon ^i,\quad \xi =\pm 1.
\end{eqnarray}
By using these Killing conditions,
the conditions $\delta\lambda^i=0$ and $\delta\psi^A=0$ 
in the SUSY transformation laws (\ref{susy_transf}) lead to
the 1/2 BPS equations 
\begin{eqnarray}
F_{\alpha \beta }&=&F_{\alpha 1}=F_{\alpha 2}=0,\quad  
{\cal D}_\alpha H^i=0,
\label{vortex_wv}\\
F_{12}&=&\xi Y_3,\quad  Y_1=Y_2=0,\quad 
 {\cal D}_1H^i-\xi i(\sigma _3)^i{}_j{\cal D}_2H^j=0.
\label{vortex_codim}
\end{eqnarray}
with $\alpha ,\beta =0,3,4,5$. 
Eq.(\ref{vortex_wv}) is satisfied by simply turning off $W_\alpha$
and dependence of $x^\alpha$.
In this section, 
we omit the directions $x^0$, $x^3$, $x^4$ and $x^5$.  

%%%%%%%%%%
{\it Vortices}:  
The remaining equation (\ref{vortex_codim}) are 
\begin{eqnarray}
F_{12}^I&=&\xi 
g^2_I\left[\zeta ^I_3-(H^{iA})^*(\sigma _3)^i{}_j(T_I)^A{}_BH^{jB}\right],
\quad Y_1=Y_2=0, \label{eq:bps_vor1}\\
0&=&{\cal D}_1H^i-\xi i(\sigma _3)^i{}_j{\cal D}_2H^j.
\label{eq:bps_vor2}
\end{eqnarray}
These equations are called vortex equations. 
They reduce to the BPS equations for 
the Abrikosov-Nielsen-Olesen (ANO) vortices~\cite{ANO}
in the case of $U(1)$ gauge theory with one charged hypermultiplet,  
whereas to those for non-Abelian vortices 
to which much attention has been recently paid 
in \cite{vortices,Eto:2005yh,Eto:2006mz,Eto:2006uw} 
in the case $U(\NC)$ gauge theory coupled to
$\NF (\geq \NC$) hypermultiplets 
in the fundamental representation with equal $U(1)$ charges.  
In order to obtain finite energy vortex solutions, we need to
turn on nonzero FI parameters $\zeta_a^I$. 
We consider the case that 
all the FI parameters take values in the third direction
as $\zeta_a^I = (0,0,\zeta^I_3)$. 
Then the condition $Y_1 = Y_2=0$ requires
that a half of the Higgs fields must vanish. 
The energy of the BPS vortices is given by 
the topological charge as
\begin{eqnarray}
 E=\xi \,{\cal V}_{12}^{(3)}.
\end{eqnarray}
Here we have defined 
\begin{eqnarray}
{\cal V}_{nm}^{(a)} \equiv
\int dx^ndx^m\left[\zeta ^I_aF_{nm}^I+2\partial _{[n}J_{m],a}\right]
= 2\pi k_I\zeta ^I_a,\quad 
 J_{n,a}\equiv i(H^{iA})^*\sigma _a\stackrel{\leftrightarrow }{\cal D}_nH^{iA},
 \label{eq:def-V}
\end{eqnarray}
with 
$k_I$ denoting the vorticity and 
$X\stackrel{\leftrightarrow }\partial Y\equiv (-\partial XY+X\partial Y)/2$.

%%%%%%%%%%
{\it Domain walls}: 
By performing the SS dimensional reduction along one of 
codimensions on the 1/2 BPS equations for vortices, 
we can derive 1/2 BPS equations for domain walls.
For example, let us perform SS dimensional reduction along the $x^2$-direction.
Then we get the following first order equations
\begin{eqnarray}
{\cal D}_1\Sigma _2&=&-\xi Y_3,\quad Y_1=Y_2=0,\label{eq:bps_wal1}\\
{\cal D}_1H^{iA}&=&\xi (\sigma _3)^i{}_j\left(\Sigma _2H^{jA}-m_{2,A}H^{jA}\right).
\label{eq:bps_wal2}
\end{eqnarray} 
We call these equations domain wall equations.
They reduce to BPS equations discussed in 
\cite{Abraham:1992vb}--\cite{IOS} 
in the case of $U(1)$ gauge theory coupled to 
$\NF$ hypermultiplets 
with equal $U(1)$ charge, whereas 
to those in \cite{INOS1}--\cite{Eto:2004vy} 
in the case $U(\NC)$ gauge theory with $\NF (>\NC)$ hypermultiplets 
in the fundamental representation. 
The topological charge of the 1/2 BPS vortices reduces to that of the
1/2 BPS walls by the SS dimensional reduction
\be
{\cal W}_{1,2}^{(3)} 
= {\cal V}_{1\check2}^{(3)}
=\int dx^1\partial _1\left[-\zeta ^I_3
\Sigma ^I_2-i(H^{iA})^*(\sigma _3\Sigma _2H)^{iA}
+i(H^{iA})^*(\sigma _3H)^{iA}m_{2,A}\right]
\ee
where we have defined 
\begin{eqnarray}
 {\cal W}_{m,n}^{(a)} \equiv {\cal V}_{m\check n}^{(a)}.
 \label{eq:def-W}
\end{eqnarray}
Then the energy of the 1/2 BPS wall is given by this charge:
\begin{eqnarray}
E=\xi \,{\cal W}_{1,2}^{(3)} .
\end{eqnarray}

\medskip 
{\it Equations in zero dimension:} 
Finally, if we fully reduce the codimensions of solitons, 
the Bogomol'nyi bound vanishes in this case of the type I\!Va 
as in the type I\!Vb. 
Therefore, spacetime-filling brane (partially
breaking SUSY vacua) cannot be
obtained as a result of the dimensional reductions. 

There exist no wavy or dyonic extensions of these solitons 
with keeping the 1/2 BPS conditions. 
Instead they become 1/4 BPS as discussed in section \ref{sec:1/4wavyanddyonic}.

%%%%%%%%%%%%%%%%%%%%%%%
\subsection{Summary of 1/2 BPS systems}
\addtocontents{toc}{\vspace{-4mm}}

We have seen that all 1/2 BPS states 
are classified into two classes, 
the types I\!Va and I\!Vb, 
with respect to unbroken SUSY. 
Unbroken SUSY in the case of two dimensional world-volume 
is ${\cal N}= (2,2)$ or $(4,0)$ for the types I\!Va or I\!Vb, respectively. 
The fundamental solitons in the type I\!Va are vortices 
whereas those in the type I\!Vb are instantons in $d=6$.  
All other solitons have been obtained by 
(SS) dimensional reductions. 
Wavy and dyonic solitons exist for the type I\!Vb. 
We summarize all descendants in Table \ref{1/2BPS-table}. 
%%%%%%%%%%%%%%%%%%%%%%%%%%%%%%%%%%
\begin{table}[ht] 
\begin{center}
\caption{\small{  
1/2 BPS states: Here, $\tilde d$ denotes spacetime dimensions 
of a soliton world-volume  
($\tilde d=1$ denotes a particle, $\tilde d=2$ a string, 
$\tilde d=3$ a membrane, and  
$\tilde d=3$ a 3-brane).
See the caption of Table~\ref{summary_table} for notations of the Roman uppercase letters. 
More notations are: w-: wavy object (or with electric flux), 
Q-: electric-charged object [that is (Q-)M means dyon].
}}
\label{1/2BPS-table} 
{\small
\begin{tabular}{|c|cc|c|c|c|c|c|} \hline
&&&$d=6$&$d=5$&$d=4$&$d=3$&$d=2$\\
&& &${\cal N}=1$ & ${\cal N}=1$ & ${\cal N}=2$&
${\cal N}=4$&${\cal N}=(4,4)$ \\ \hline\hline
\text{type I\!Vb}&$\tilde d=2,$&$ {\cal N}=(4,0)$&
 (w-)I&(w-)M&(w-)H&(w-)N& wave \\
&$\tilde d=1$,&${\cal N}=4$ & -&(Q-)I& (Q-)M&(Q-)H&(Q-)N \\ \hline
&$\tilde d=4$,&${\cal N}=1$ & V&W&-&-&-\\
\text{type I\!Va}&$\tilde d=3$,&${\cal N}=2$ & -&V&W&-&-\\
&$\tilde d=2$,&$ {\cal N}=(2,2)$ & -&-&V&W&-\\
&$\tilde d=1$,&${\cal N}=4$ & -&-&-&V&W\\
\hline
\end{tabular}
}
\end{center}
\end{table}

The types I\!Va and I\!Vb exhibit very different features. 
Since BPS solitons in the types I\!Vb and I\!Va 
preserve ${\cal N}=(4,0)$ and $(2,2)$ SUSY, respectively,  
in the case of two dimensional world volume, 
the moduli space for these solitons 
are described by ${\cal N} = (4,0)$ nonlinear sigma models 
on hyper-K\"ahler manifolds (with torsion)~\cite{Howe:1987qv} and 
${\cal N} = (2,2)$ nonlinear sigma models 
on K\"ahler manifolds~\cite{Zumino:1979et}, 
respectively. 
Second, BPS solitons in the type I\!Vb naturally 
exist in theories with sixteen supercharges 
(typically in $d=4$, ${\cal N}=4$ SUSY Yang-Mills theory) 
whereas BPS solitons in the type I\!Va 
can exist in theories with 
four supercharges ($d=4$, ${\cal N}=1$ SUSY theory).

We would like to note that 
there exist the two kinds of objects in 
co-dimensions two and one. 
The Hitchin vortices (type I\!Vb) and 
the ordinary vortices of the ANO type (type I\!Va) 
are with codimension two, whereas 
the Nahm walls (type I\!Vb) and 
the ordinary walls (type I\!Va) 
are with codimension one.
The Hitchin vortices and 
the Nahm walls live inside the unbroken (non-Abelian) phase
whereas ordinary vortices and walls 
live inside the Higgs phase. 
This distinction becomes very important 
when we consider 1/4 and 1/8 BPS composite states 
in the following sections. 
As denoted in the text there exist no finite energy solutions 
for Hitchin vortices or Nahm walls in non-compact spaces.

Before closing this section we make comments 
on solutions known so far. 
The method to construct multi-instantons and multi-BPS monopoles 
were established by Atiyah, Hitchin, Drinfeld
and Manin (ADHM) \cite{Atiyah:1978ri}  
and Nahm \cite{Nahm}, respectively. 
Their moduli spaces are both 
hyper-K\"ahler manifolds. 
The ${\cal N}= (4,0)$ nonlinear sigma model  
on the ADHM moduli space was considered in \cite{Witten:1994tz}. 
These BPS instantons can be realized
by D$p$-branes on D$(p+4)$-branes~\cite{Wi} 
whereas BPS monopoles by D$p$-branes ending on (stretched between) 
D$(p+2)$-branes~\cite{brane-monopole}
in type I\!IA/I\!IB superstring theories. 
These brane configurations give clear physical interpretations of
the ADHM/Nahm constraints as the F- and D- flatness conditions 
in the SUSY gauge theory on D$p$-brane world-volume. 
On the other hand 
1/2 BPS equations in the type I\!Va 
can be (partially) solved by the method of 
the moduli matrix (walls in \cite{INOS1,Eto:2005wf} 
and vortices in \cite{Eto:2005yh,Eto:2006mz,Eto:2006pg}). 
Fully exact solutions can be obtained 
in the strong gauge coupling limit in which 
the model reduces to a hyper-K\"ahler nonlinear sigma model.
Some brane configurations are also available; 
vortices by Hanany and Tong~\cite{vortices} 
and walls in \cite{Lambert:1999ix,Eto:2004vy,Hanany:2005bq}.

%%%%%%%%%%%%%%%%%%%%%%%%%%%%%%%%%%%%%%%%%%%%%%%%%%%%%%%%%%%%%%%%%%
\section{1/4 BPS Systems} \label{sect.4}
%%%%%%%%%%%%%%%%%%%%%%%%%%%%%%%%%%%%%%%%%%%%%%%%%%%%%%%%%%%%%%%%%%
\addtocontents{toc}{\vspace{-2.3mm}}

In the previous section we have discussed different 1/2 BPS solitons; 
instanton, monopole, vortex and domain wall 
including their wavy and dyonic extension.
In this section we will present several 1/4 BPS equations 
for composite states of those 1/2 BPS solitons. 
As we have mentioned in Sec.\ref{sect.2}, 
there exist two sets of 1/4 SUSY projection operators, 
the type I\!Ia in Eq.(\ref{type A}) and 
the type I\!Ib in Eq.(\ref{type B}).
Similarly to the method in Sec.\ref{sect.3}, 
we will start with the
1/4 BPS equations in 6 dimensions and derive diverse 1/4 BPS equations
in $d=5,4,3$ by the SS dimensional reductions.
These 1/4 BPS states certainly involve the 1/2 BPS vortices and/or
1/2 BPS domain walls which are typical solitons in the Higgs phase
of supersymmetric gauge theories.
Therefore, the Higgs fields play important roles in the following.

%%%%%%%%%%%%%%%%%%%%%%%%%%%%%%%%%%%%%%%%%%%%%%%%%%%%%%%%%%%%%%%%%%
\subsection{Series of the type I\!Ib}  \label{Sec.IIb} 
%%%%%%%%%%%%%%%%%%%%%%%%%%%%%%%%%%%%%%%%%%%%%%%%%%%%%%%%%%%%%%%%%%
\addtocontents{toc}{\vspace{-2.3mm}}
In this section, 
subsections \ref{sec:ivv}, \ref{sec:mvw} and \ref{sec.junction} 
are devoted to a review whereas 
subsections  \ref{HVNW} and \ref{sec:1/4wavyanddyonic} 
are new contributions. 

%%%%%%%%%%%%%%%%%%%%%%%%%%%%%%%%%%%%%%%%%%%%%%%%%%%%%%%%%%%%%%%%%%
\subsubsection{Instanton - Vortex}  \label{sec:ivv}
%%%%%%%%%%%%%%%%%%%%%%%%%%%%%%%%%%%%%%%%%%%%%%%%%%%%%%%%%%%%%%%%%%
\addtocontents{toc}{\vspace{-2.3mm}}

We begin with the type I\!Ib 1/4 SUSY projections in 6 dimensions
\begin{eqnarray}
\gamma ^{05}\varepsilon ^i=\eta \varepsilon ^i, \quad 
\gamma ^{12}\left(i\sigma _3\varepsilon\right) ^i=\xi \varepsilon ^i,
\quad \gamma ^{34}\left(i\sigma _3\varepsilon\right) ^i=\xi '\varepsilon^i
\end{eqnarray}
where $\eta,\xi,\xi'$ are $\pm 1$ satisfying $\eta\xi\xi' = -1$.
Imposing invariance to these 1/4 SUSY as $\delta\lambda^i=0$ and
$\delta\psi^A = 0$ given in Eq.(\ref{susy_transf}), 
we can derive a set of 1/4 BPS equations \cite{Eto:2004rz} 
\begin{eqnarray}
&&F_{13}+\eta F_{24}=0,\quad F_{14}-\eta F_{23}=0,\quad
\xi F_{12}+\xi 'F_{34}=Y_3,\quad Y_1=Y_2=0,
\label{ivv1}\\
 &&{\cal D}_1H^i-\xi i(\sigma _3)^i{}_j{\cal D}_2H^j=0,\quad 
{\cal D}_3H^i-\xi 'i(\sigma _3)^i{}_j{\cal D}_4H^j=0.
\label{ivv2}
\end{eqnarray}
We may call these equations as the SDYM-Higgs equations. 
They comprise 
the (anti) SDYM equations for instantons localized in
the $x^1$-$x^2$-$x^3$-$x^4$ space, 
and the BPS equations for the vortices
living in the $x^1$-$x^2$ plane and in the $x^3$-$x^4$ plane. 
The world-volume of these solitons can be 
summarized as follows.
\begin{eqnarray*}
\begin{array}{c|cccccc|cc}
{\rm IVV}&0 &1 &2 &3 &4 &(5) &  &{\rm sign}\\
\hline
{\rm Instanton}&\bigcirc &\times &\times &\times &\times &\bigcirc &{\bf 1}_2&\eta \\
{\rm Vortex}&\bigcirc &\times &\times &\bigcirc &\bigcirc &\bigcirc &i\sigma _3 &\xi \\
{\rm Vortex}&\bigcirc &\bigcirc &\bigcirc &\times &\times &\bigcirc &i\sigma _3&\xi ' 
\end{array}
\end{eqnarray*}
Here ``$\bigcirc$" denotes the world-volume directions 
of the solitons, whereas ``$\times$" denotes their codimensional directions.
From the above table we find that 
this composite soliton trivially extends to the $x^5$-direction and 
therefore that the dimensionally reduced 
configuration exists in 5 dimensions also. 
The energy of this composite solitons can be written by
the sum of the three topological charges as
\begin{eqnarray}
 E_{\rm m}=\eta {\cal I}_{1234}+ \left(\xi  {\cal V}_{12}^{(3)}\right)v_{34}
+ \left(\xi' {\cal V}_{34}^{(3)}\right)v_{12} , 
\end{eqnarray}
where we have ignored the sixth direction $x^5$.  
Here ${\cal I}_{1234}$ is the instanton charge (\ref{eq:def-I}),  
and ${\cal V}_{mn}^{(3)}$'s are the vortex charge (\ref{eq:def-V}),
and 
the quantities $v_{34}$ and $v_{12}$ denote 
the world-volumes of vortices along $x^3$-$x^4$ and 
$x^1$-$x^2$, respectively. 

The equations (\ref{ivv1}) and (\ref{ivv2}) were firstly found 
in general K\"ahler manifold as a base space 
by mathematicians \cite{math},  
without the BPS property or SUSY. 
There they are simply called vortex equations 
although they contain instantons also.
These equations are rediscovered 
by Hanany and Tong in \cite{vm} in the case 
of $U(\NC)$ gauge theory coupled to $\NF$ hypermultiplets 
in the fundamental representation with equal $U(1)$ charges.
We showed in \cite{Eto:2004rz} that 
these equations are 1/4 BPS and then 
studied this system with $\NF = \NC$ 
in 5 dimensions in detail.
In the paper \cite{Eto:2004rz} we further 
found that the 1/4 BPS equations (\ref{ivv1}) and (\ref{ivv2}) 
describe two different kinds of composite states, 
1) instantons in the Higgs phase and 
2) intersections of 
vortices in the $x^1$-$x^2$ and the $x^3$-$x^4$ planes.

1)
As the former concerns, 
when we put ordinary instantons into the Higgs phase by turning on 
the FI-parameters, 
they can be stabilized if the magnetic flux is squeezed to form a vortex 
similarly to the Meissner effect of 
the monopoles in the Higgs phase. 
The total system becomes instantons lying inside 
vortex-3-branes (membranes) in $d=6$ ($d=5$). 
The instanton in the Higgs phase has  
the same mass $8\pi^2/g^2$
with that of the ordinary instanton.\footnote{
These instantons can be regarded as lumps 
in the ${\bf C}P^{N-1}$ model as 
the effective field theory on the vortex 
in the case of $U(N)$ gauge theory with 
$N$ hypermultiplets in the fundamental representation~\cite{Eto:2004rz}.
}

2)
In the latter case, the 1/4 BPS equations (\ref{ivv1}) and (\ref{ivv2}) 
describe a very different situation from the instantons 
in the Higgs phase \cite{Eto:2004rz}.
It can exist even for the Abelian gauge theory, 
contrary to the ordinary instantons which 
do not exist due to the trivial instanton charge $\pi_3 = 0$.
Let us consider the intersection of 
vortices with vorticity $k$ in the $x^1$-$x^2$ plane and 
vortices with the vorticity $k'$ in the $x^3$-$x^4$ plane. 
Interestingly, the contribution of the instanton charge 
to the energy is nonzero for intersecting vortices.
It can be easily shown as follows.
First we recall
$\xi\int dx^1dx^2 F_{12} = 2\pi \xi k>0$ and
$\xi'\int dx^3dx^4 F_{34} = 2\pi \xi'k'>0$.
Then the instanton contribution can be calculated as 
$E_{\rm m} = \eta {\cal I}_{1234} = (2\eta/g^2) 
\int dx^4 F_{12}F_{34} = (\eta kk')8\pi^2/g^2 $.
Due to the relation $\eta\xi\xi'=-1$, 
the contribution is always negative
and is proportional to the product of 
the vorticities $E_{\rm m} = - |kk'|8\pi^2/g^2 $.
Such objects with the negative energy should not 
be regarded as solitons,
but they are rather binding energy for the intersections of the solitons.

Thus the 1/4 BPS equations (\ref{ivv1}) and (\ref{ivv2}) describe
both instantons in the Higgs phase and 
intersections of the vortices. 
As we will see in the following subsections, 
the monopole charge and the Hitchin charge, 
derived from the instanton charge by dimensional reductions,
also can become either positive energies of ordinary solitons 
or negative binding energies localized on intersections of solitons,  
depending on configurations.

%%%%%%%%%%%%%%%%%%%%%%%%%%%%%%%%%%%%%%%%%%%%%%%%%%%%%%%%%%%%%%%%%%
\subsubsection{Monopole - Vortex - Wall}  \label{sec:mvw}
%%%%%%%%%%%%%%%%%%%%%%%%%%%%%%%%%%%%%%%%%%%%%%%%%%%%%%%%%%%%%%%%%%
\addtocontents{toc}{\vspace{-2.3mm}}

As was shown in the Sec.\ref{sect.3}, 
the 1/2 BPS equations for instantons and 
vortices reduce to those for 
monopoles and domain walls, respectively, 
by performing the SS dimensional reduction 
along one of their codimensions.
When we perform the SS dimensional reduction 
(along, for example, the $x^4$-direction) 
in the above 1/4 BPS eqs.~(\ref{ivv1}) and (\ref{ivv2}) for
composite states of instantons, vortices in the $x^1$-$x^2$ plane and 
vortices in the $x^3$-$x^4$ plane, 
we derive another set of 1/4 BPS equations 
for composite states of monopoles 
in the $x^1$-$x^2$-$x^3$ space, vortices in the $x^1$-$x^2$ plane
and domain walls perpendicular to the $x^3$-coordinate, given by  
\begin{eqnarray}
F_{13}-\eta {\cal D}_2\Sigma _4=0,\quad 
{\cal D}_1\Sigma _4+\eta F_{23}=0,\quad
\xi F_{12}-\xi '{\cal D}_3\Sigma _{4}=Y_3,\quad
Y_1=Y_2=0,   \label{wvm1} \\
{\cal D}_1H^{iA}-\xi i(\sigma _3)^i{}_j{\cal D}_2H^{jA}=0,\quad 
{\cal D}_3H^{iA} - \xi '(\sigma _3)^i{}_j
\left[(\Sigma _4)^A{}_BH^{jB}-m_{4,A}H^{jA}\right]=0.  \label{wvm2} 
\end{eqnarray}
The codimensions and the world-volume directions 
of solitons in this 1/4 BPS system 
is summarized as follows.
\begin{eqnarray*}
\begin{array}{c|ccccc|cc}
{\rm MVW}&0 &1 &2 &3 &(5) &  &{\rm sign}\\
\hline
{\rm Monopole}&\bigcirc &\times &\times &\times &\bigcirc &{\bf 1}_2&\eta \\
{\rm Vortex}&\bigcirc &\times &\times &\bigcirc &\bigcirc &i\sigma _3 &\xi \\
{\rm Wall}&\bigcirc &\bigcirc &\bigcirc &\times &\bigcirc &i\sigma _3&\xi '
\end{array}
\end{eqnarray*}
This 1/4 BPS system trivially depends on the $x^5$-direction,
so that these composite solitons exist in four dimensional spacetime. 
The energy of this composite solitons is three sum of the 
topological charges of the monopoles, vortices and walls
\begin{eqnarray}
E_{\rm m}=\eta {\cal M}_{123}+ \left(\xi {\cal V}_{12}^{(3)}\right)v_3
+ \left(\xi ' {\cal W}_{3,4}^{(3)}\right)v_{12},
\end{eqnarray}
with the monopole charge (\ref{eq:def-M}), 
the vortex charge (\ref{eq:def-V}),
and the domain wall charge (\ref{eq:def-W}).

The 1/4 BPS equations (\ref{wvm1}) and (\ref{wvm2}) (without walls) 
were firstly found by Tong in \cite{vm} in the case 
of $d=4$, ${\cal N}=2$ SUSY $U(\NC)$ gauge theory 
with $\NF = \NC$ hypermultiplets 
in the fundamental representation.
Similarly to the 1/4 BPS states of the instantons, 
vortices and vortices discussed in the last subsection, 
the monopole charge in this 1/4 BPS system also exhibits two different features.  
One is a monopole in the Higgs phase~\cite{vm}.
If an ordinary monopole of the non-Abelian gauge theory
in the Coulomb phase is put into the Higgs phase, 
magnetic fluxes from it are squeezed due to the Meissner effect 
to form vortices, 
making the total system a composite state of 
a monopole attached by vortices.\footnote{
These monopoles can be regarded as kinks 
in the ${\bf C}P^{N-1}$ model with a potential 
as the effective field theory on a vortex 
in the case of $U(N)$ gauge theory with 
massive $N$ hypermultiplets in the fundamental representation, 
as found by Tong~\cite{vm}.
} 
In that situation the mass of the monopole is the same with 
the mass $4\pi|\Delta\langle \Sigma_4 \rangle|/g^2$ of 
an ordinary monopole in the Coulomb phase.  
On the other hand, if we consider $\NF (> \NC)$ hypermultiplets 
domain walls are also allowed~\cite{INOS3}. 
In the Abelian gauge theory the 1/4 BPS equation (\ref{wvm1}) and (\ref{wvm2})
describe the composite state of the vortices ending on the walls
rather than the monopoles in the Higgs phase.
Interestingly, the monopole charge takes a nonzero value 
even for the Abelian gauge theory. 
In fact it gives the negative energy
contribution $-4k\pi|\Delta\langle \Sigma_4 \rangle|/g^2$ 
when $k$ ANO vortices penetrate a wall. 
A vortex is divided by the wall into two vortices 
attached to one and the other sides of the wall, 
and end points of these two vortices 
can be separated along the wall world-volume.\footnote{
This is a peculiar situation for this system. 
In the case of vortex intersections discussed in 
the previous section one vortex cannot be divided by the other vortex 
from their dimensionality. 
}  
Each has the half $- 2\pi|\Delta\langle \Sigma_4 \rangle|/g^2$ 
of the (negative) unit monopole charge. 
Although these negative energies are localized at the ending points
of the vortices on the wall, they should not be regarded as solitons.
Rather they are binding energy of the junction 
called boojums~\cite{INOS3,Sakai:2005sp}. 
We constructed implicit solutions 
for this system in $d=5$, ${\cal N}=1$ gauge theory. 
Moreover we have obtained all the exact solutions 
in the strong gauge coupling limit \cite{INOS3}.  
Now we stand at the position 
where the negative binding energy 
of the boojum can be understood 
from the viewpoint of the
SS dimensional reduction.
First recall that the instanton charge ${\cal I}_{mnkl}$ 
reduces to the monopole charge 
${\cal M}_{mnk} = {\cal I}_{mnk\check l}$. 
This relation is kept even when instantons and monopoles 
are put into the Higgs phase.
Furthermore, this relation is also correct for the charges 
as the binding energy
for intersections of vortices and vortices ending on walls.
When we push to perform the SS dimensional reduction once more as we will
see below soon, very similar discussion will be able to be done again.

We do not know if there exist 
composite states made of only monopoles and walls without
vortices, but if they exist they are 1/4 BPS. 
However they are likely impossible because of the Meissner effect.

%%%%%%%%%%%%%%%%%%%%%%%%%%%%%%%%%%%%%%%%%%%%%%%%%%%%%%%%%%%%%%%%%%
\subsubsection{Hitchin vortex -  Wall \ \ (Domain wall junction)}  \label{sec.junction} 
%%%%%%%%%%%%%%%%%%%%%%%%%%%%%%%%%%%%%%%%%%%%%%%%%%%%%%%%%%%%%%%%%%
\addtocontents{toc}{\vspace{-2.3mm}}

As we have seen in the Sec.\ref{sect.3},
the 1/2 BPS equations of monopoles reduce to 
those for the Hitchin system by one dimensional reduction. 
Vortices reduce to walls by the SS dimensional reduction.
Therefore, performing the SS dimensional reduction 
in Eqs.~(\ref{wvm1}) and (\ref{wvm2}) along $x^2$,  
we obtain the following 1/4 BPS equations 
for composite states of Hitchin vortices 
in the $x^3$-$x^1$ plane, 
walls perpendicular to the $x^1$-direction and 
walls perpendicular to the $x^3$-direction, given by 
\begin{eqnarray}
&&F_{13}+\eta i[\Sigma _2,\Sigma _4]=0,\quad
{\cal D}_1\Sigma _4+\eta {\cal D}_3\Sigma _2=0,
\label{jww1}\\
&&\xi {\cal D}_1\Sigma _2+\xi '{\cal D}_3\Sigma _4=-Y_3,\quad Y_1=Y_2=0,
\label{jww2}\\
&&{\cal D}_1H^{iA}-\xi (\sigma _3)^i{}_j
\left[(\Sigma _2)^A{}_BH^{jB}-m_{2,A}H^{jA}\right]=0,
\label{jww3}\\
&&{\cal D}_3H^{iA}-\xi '(\sigma _3)^i{}_j
\left[(\Sigma _4)^A{}_BH^{jB}-m_{4,A}H^{jA}\right]=0. 
\label{jww4}
\end{eqnarray} 
The configuration can be summarized as follows. 
\begin{eqnarray*}
\begin{array}{c|cccc|cc}
{\rm HWW}&0 &1 &3 &(5) &  &{\rm sign}\\ \hline
{\rm Hitchin\ vortex}&\bigcirc &\times &\times &\bigcirc &{\bf 1}_2&\eta \\
{\rm Wall}&\bigcirc &\times &\bigcirc &\bigcirc &i\sigma _3 &\xi \\
{\rm Wall}&\bigcirc &\bigcirc &\times &\bigcirc &i\sigma _3 &\xi '   
\end{array}
\end{eqnarray*}  
The energy of this 1/4 BPS states is given by 
\begin{eqnarray}
 E_{\rm m}=\eta {\cal H}_{13} + \left(\xi {\cal W}_{1,2}^{(3)}\right)v_3
+ \left(\xi '{\cal W}_{3,4}^{(3)}\right)v_1 \;.
\end{eqnarray}

Recently, the 1/4 BPS equations (\ref{jww1}), (\ref{jww2}),
(\ref{jww3}) and (\ref{jww4}) 
have been found \cite{Eto:2005cp} 
in the case of $d=4$, ${\cal N}=2$ 
$U(\NC)$ gauge theory coupled to 
$\NF$ hypermultiplets with complex masses
in the fundamental representation 
with equal $U(1)$ charges. 
All possible moduli have been determined, 
and all the exact solutions have been obtained 
in the strong gauge coupling limit. 
Especially,   
the 1/4 BPS states in the Abelian gauge theory have been 
extensively studied, and it has been found that 
several walls can constitute webs of walls, 
like $(p,q)$ string/5-brane webs in superstring theory. 
In the Abelian gauge theory, 
the charge for the Hitchin vortices 
does not give positive contributions to the energy. 
Instead, it gives negative binding energy localized 
at the junction lines (points) of the walls in $d=3$ ($d=4$). 
There were many works on the 1/4 BPS domain wall junctions 
in theories with both eight supercharges and four supercharges.
All exact solutions of the wall junctions known so far 
carry negative junction charges 
as binding energy~\cite{Oda:1999az,Eto:2005cp}. 
The negativeness of the Hitchin charge as binding energy 
of the domain wall junctions can be understood 
by remembering that it 
is obtained by the SS dimensional reduction(s) 
of the monopole charge for 
the boojums at the end points of vortices on the wall  
(or of the negative instanton charge localized 
at the vortex intersections).  

On the other hand, 
we have seen in the last two subsections 
that there exist 1/4 BPS states 
with positive energy of instantons or monopoles, 
namely instantons or monopoles in the Higgs phase, respectively. 
This observation strongly suggests 
that there should exist 
the 1/4 BPS composite states 
of the Hitchin vortices with positive energy, 
which are put into the Higgs phase by the FI-term. 
In fact we found in \cite{Eto:2005fm} that 
the Hitchin vortices 
with positive energy live 
on the junction lines (points) of walls 
in the non-Abelian gauge theories. 
Generally, walls in the non-Abelian gauge theories
can constitute very ample webs, 
containing both negative and positive Hitchin charges. 
Therefore wall webs in the non-Abelian gauge theory 
have a rich structure 
which does not exist 
in the Abelian gauge theory~\cite{Eto:2005fm}.

%%%%%%%%%%%%%%%%%%%%%%%%%%%%%%%%%%%%%%%%%%%%%%%%%%%%%%%%%%%%%%%%%%%%
\subsubsection{Hitchin vortex - Vortex, and  Nahm wall - Wall}
 \label{HVNW}
%%%%%%%%%%%%%%%%%%%%%%%%%%%%%%%%%%%%%%%%%%%%%%%%%%%%%%%%%%%%%%%%%%%%
\addtocontents{toc}{\vspace{-2.3mm}}

If we perform the SS dimensional reduction 
in Eqs.~(\ref{wvm1}) and (\ref{wvm2}) 
along the $x^3$-direction instead of the $x^2$-direction, 
we obtain a set of 1/4 BPS equations which include  
the 1/2 BPS equations for the Hitchin vortices and 
for ordinary vortices, where both objects would 
lie perpendicular to the $x^1$-$x^2$ plane.
By taking a further SS dimensional reduction 
along the $x^2$-direction 
after the above operation, 
we obtain the equations containing   
the 1/2 BPS equations for the Nahm  walls and for ordinary walls, 
where both objects would 
be perpendicular to the $x^1$-direction.  
These objects could lie as the following. 
\begin{eqnarray*}
\begin{array}{c|cccc}
{\rm HV}&0 &1 &2 &(5) \\ \hline
{\rm Hitchin\ vortex}&\bigcirc &\times &\times &\bigcirc  \\
{\rm vortex}&\bigcirc &\times &\times &\bigcirc  \\
\end{array}
{},\quad 
\begin{array}{c|ccc}
{\rm NW}&0 &1 &(5) \\ \hline
{\rm Nahm\ wall}&\bigcirc &\times &\bigcirc  \\
{\rm wall}&\bigcirc &\times &\bigcirc \\
\end{array} 
\end{eqnarray*}
It seems, however, quite difficult for these objects,
a Hitchin vortex and an ordinary vortex, 
or a Nahm wall and an ordinary wall 
to coexist. 
This is because the Hitchin vortices and the Nahm walls 
exist in an unbroken phase of non-Abelian gauge group, 
whereas ordinary walls and vortices exist 
in the Higgs phase and 
their tension is proportional to the FI-term 
of the Abelian gauge group. 
As a further negative evidence, 
there exist the condition,
\begin{eqnarray}
 \Phi H^1=\Phi^\dagger H^2=0,\quad  (\Phi)^A{}_B=(\Sigma_3-i\xi'\Sigma_4)^A{}_B
-(m_{3,A}-i\xi'm_{4,A})\delta^A{}_B
\end{eqnarray} 
as a part of the reduced 1/4 BPS equations, 
implying that the existence of 
one excludes the other. 
Therefore, it is likely the case that 
this 1/4 BPS system contains
only one kind of these BPS solitons 
and then it reduces to a 1/2 BPS state.

%%%%%%%%%%%%%%%%%%%%%%%%%%%%%%%%%%%%%%%%%%%%%%%%%%%%%%%%%%%%%%%%%%
\subsubsection{Wavy and dyonic extension} \label{sec:1/4wavyanddyonic}
%%%%%%%%%%%%%%%%%%%%%%%%%%%%%%%%%%%%%%%%%%%%%%%%%%%%%%%%%%%%%%%%%%
\addtocontents{toc}{\vspace{-2.3mm}}

Similarly to 1/2 BPS equations of the type I\!Vb,
the 1/4 BPS equations of the type I\!Ib 
can be extended to wavy or dyonic objects 
by turning on the $x^0$ and $x^5$ dependence, 
since both projection operators have $\gamma^{05}$.

{\it Wavy solitons}: 
In addition to the 1/4 BPS equations (\ref{ivv1}) and (\ref{ivv2})
and their descendants given by the SS dimensional reductions,  
taken ($6-d$)-times along the $x^p$-directions $(p=d-1,\cdots,4)$,
1/4 BPS equations for wavy extensions are obtained, to yield 
\begin{eqnarray}
 F_{0n}+\eta F_{5n}=0,\quad 
  \D_0\Sigma_p + \eta \D_5\Sigma_p = 0,\quad
 F_{05}=0,\quad 
  \left({\cal D}_0+\eta {\cal D}_5\right)H^i=0,
 \label{wavy_1/4}
\end{eqnarray}
with $n=1,\cdots,d-2$.
These reduce to  the same simple equations with 
Eq.(\ref{eq_05_5}) with appropriate gauge fixing 
under a given static background.
We obtain their wavy extension 
by promoting any moduli parameters $\phi^i$ of 
that background
to arbitrary functions $\phi^i(x^0-\eta x^5)$.
The solutions have to satisfy the Gauss law
\be
\frac{1}{g_I^2}\D_nF_{0n}^I 
+i\sum_{p=d-1}^4\left[\Sigma_p,\D_0\Sigma_p\right]^I
- 2i(T^I)^A{_B}H^{iB}\overset{\leftrightarrow}{\D}_0(H^{iA})^* = 0
\ee
in addition to the BPS equations. 
The energy is given by sum of the energies of the solitons and 
the fifth component of the Poynting vector, 
$E = E_m -\eta S_5'$ with
\be
S_5' = \frac{1}{g_I^2} \int d^{d-1}x
\left(
F^I_{0n}F^I_{5n} + \sum_{p=d-1}^4\D_0\Sigma^I_p\D_5\Sigma^I_p
+ (\D_0H^{iA})^*(\D_5H^{iA}) + (\D_5H^{iA})^*(\D_0H^{iA})
\right).
\label{S5'}
\ee

{\it Dyonic solitons}:
When we perform the trivial dimensional reduction along the $x^5$-direction,
1/4 BPS wavy solitons reduce to 1/4 BPS dyonic solitons.
By this dimensional reduction, Eqs.~(\ref{wavy_1/4}) reduce to
\be
F_{0n}+\eta \D_n\Sigma_5 = 0,\ 
\D_0\Sigma_p + i\eta\left[\Sigma_5,\Sigma_p\right] = 0,\ 
\D_0\Sigma_5 = 0,\ 
\D_0 H^{iA} - i\eta (\Sigma_5)^A{_B} H^{iB} = 0.
\label{dyon_1/4}
\ee
These can be solved by the solution given in Eq.(\ref{eq_05_5}).
Combining Eqs.~(\ref{dyon_1/4}), the Gauss law 
\be
\frac{1}{g_I^2}\left(
\D_nF_{0n}^I + i\sum_{p=d-1}^4 \left[\Sigma_p,\D_0\Sigma_p\right]^I
\right) - 2i(T^I)^A{_B}H^{iB}\overset{\leftrightarrow}{\D}_0(H^{iA})^* = 0 
\ee
determines the configuration of the $\Sigma_5$. 
The fifth component of the Poynting vector given in Eq.(\ref{S5'})
reduces to the electric charge
\be
 S_5' \to 
 Q_{\rm e} 
&=&  \frac{1}{g_I^2} \int d^{d-1}x
\bigg(
F^I_{0n}\D_n\Sigma_5^I + i\sum_{p=d-1}^4\D_0\Sigma^I_p\left[\Sigma_p,\Sigma_5\right]^I
  \nn\\ 
&& - i (\D_0H^{iA})^*(\Sigma_5H^{iA}) + i(\Sigma_5H^{iA})^{*}(\D_0H^{iA})
 \bigg)\nonumber\\
&=& \frac{1}{g_I^2}\int d^{d-1}x\ \partial_n\left(\Sigma_5^IF^I_{0n}\right)
  \label{S5''}
\ee
and the energy is given by $E = -\eta Q_{\rm e} + E_{\rm m} (\ge 0)$.

%%%%%%%%%%%%%%%%%%%%%%%%%%%%%%%%%%%%%%%%%%%%%%%%%%%%%%%%%%%%%%%%%%
\subsection{Series of the type I\!Ia}    \label{Sec.IIa} 
%%%%%%%%%%%%%%%%%%%%%%%%%%%%%%%%%%%%%%%%%%%%%%%%%%%%%%%%%%%%%%%%%%
\addtocontents{toc}{\vspace{-2.3mm}}

While the type I\!Ib series of the 1/4 BPS equations 
in the previous subsection 
has been found recently and studied so far, 
the most of 1/4 BPS equations of the type I\!Ia
given in this subsection are unknown and new. 
We do not make an effort to solve the BPS equations in this paper. 
Instead we concentrate on finding new 1/4 BPS equations, 
classifying them 
and discussing their physical properties.
We leave detailed analysis of these 1/4 BPS equations as 
future problems.

%%%%%%%%%%%%%%%%%%%%%%%%%%%%%%%%%%%%%%%%%%%%%%%%%%%%%%%%%%%%%%%%%%
\subsubsection{Vortices}
%%%%%%%%%%%%%%%%%%%%%%%%%%%%%%%%%%%%%%%%%%%%%%%%%%%%%%%%%%%%%%%%%%
\addtocontents{toc}{\vspace{-2.3mm}}

Let us turn our attention to the other 1/4 SUSY projection in 6 dimensions.
We consider the following type I\!Ia operators
\begin{eqnarray}
\gamma ^{23}(i\sigma _1 \varepsilon )^i =\xi_1 \varepsilon ^i,\quad 
\gamma ^{31}(i\sigma _2 \varepsilon )^i =\xi_2 \varepsilon ^i,\quad 
\gamma ^{12}(i\sigma _3 \varepsilon )^i =\xi_3 \varepsilon ^i,\quad 
\xi _1\xi _2\xi _3=-1.
  \label{pj_vvv}
\end{eqnarray}
By imposing SUSY preservation 
$\delta\lambda^i=0$ and $\delta\psi^A = 0$ 
in the SUSY transformation laws (\ref{susy_transf})
with the Killing conditions (\ref{pj_vvv}),
we find new 1/4 BPS equations
\begin{eqnarray}
&&F_{\alpha \beta }=F_{\alpha n}=0,\quad
{\cal D}_\alpha H^i=0,
\label{vvv1}\\
&& \sum_{n=1}^3(\tau_n)^i{_j}{\cal D}_nH^j=0,\quad 
 \myfrac12 \epsilon _{nml}F_{ml}=\xi _nY_n ,
 \label{vvv2}
\end{eqnarray}
where we have defined $\tau _n\equiv \xi _n i\sigma _n$ and 
$\alpha ,\beta =0,4,5$ and $n,m,l=1,2,3$.
Equation (\ref{vvv1}) is 
satisfied by simply ignoring the $x^\alpha$ dependence
and by ignoring $W_\alpha$.
Equation (\ref{vvv2}) consists of 
three 1/2 BPS equations
describing vortices in the $x^1$-$x^2$ plane, 
those in the $x^2$-$x^3$-plane and those in the $x^3$-$x^1$-plane.
The configuration can be summarized as follows.
\begin{eqnarray*}
\begin{array}{c|cccccc|ccc}
{\rm VVV}&0 &1 &2 &3 &(4) &(5) & & {\rm FI} &{\rm sign}\\ \hline
{\rm Vortex}&\bigcirc &\bigcirc &\times &\times 
&\bigcirc &\bigcirc &i\sigma _1& \zeta_1^I & \xi _1\\
{\rm Vortex}&\bigcirc &\times &\bigcirc &\times 
&\bigcirc &\bigcirc &i\sigma _2& \zeta_2^J & \xi _2\\
{\rm Vortex}&\bigcirc &\times &\times &\bigcirc 
&\bigcirc &\bigcirc &i\sigma _3& \zeta_3^K & \xi _3
\end{array}
\end{eqnarray*} 
Here $\zeta_1^I$, $\zeta_2^J$, $\zeta_3^K$ 
($I\neq J \neq K \neq I$)
denote first, second, third components of 
three different triplets of the FI-parameters, 
respectively, as explained below. 
The energy of these composite states 
is the sum of the energy of 
vortices in the three directions, given by 
\begin{eqnarray}
 E_{\rm m}=
\left(\xi _3{\cal V}_{12}^{(3)}\right)v_3
+\left(\xi _1{\cal V}_{23}^{(1)}\right)v_1
+\left(\xi _2{\cal V}_{31}^{(2)}\right)v_2. \label{vvv-energy}
\end{eqnarray}

In the case that the moduli space of vacua is 
a non-trivial Higgs branch ${\cal M}_{\rm vac.}$,  
the model reduces to the hyper-K\"ahler nonlinear sigma model 
on ${\cal M}_{\rm vac.}$ in the strong gauge coupling limit. 
Vortices reduce to lumps and 
the equations (\ref{vvv1}) and (\ref{vvv2}) 
reduce to those for lump intersections 
found in \cite{Naganuma:2001pu,Portugues:2002ih}. 
(The Killing conditions (\ref{pj_vvv}) were already 
found in \cite{Naganuma:2001pu}.)
The energy bound (\ref{vvv-energy}) reduces to the sum of 
the pull-back of three complex structures 
${\bf I}$, ${\bf J}$ and ${\bf K}$
on the hyper-K\"ahler manifold ${\cal M}_{\rm vac.}$ 
to the $x^1$-$x^2$, $x^2$-$x^3$ and $x^3$-$x^1$ planes, respectively.   

We need to turn on an FI parameter 
to obtain a vortex solution. 
The direction of the FI parameters 
$(\zeta_1^I,\zeta_2^I,\zeta_3^I)$ 
in the three dimensional space 
has to be parallel to the direction of 
the unit vector $(n_1,n_2,n_3)$ 
in the 1/2 SUSY projection operator $\gamma^{nm}(\vec n\cdot i\vec\sigma)$
for the vortices in the $x^n$-$x^m$ plane \cite{Naganuma:2001pu}.  
Therefore we need to introduce several (at least three)
FI parameters in different directions 
with gauge group containing several $U(1)$ factors,  
to obtain intersecting vortices.
Such FI-parameters were summarized in the table above. 

We do not know if there exist nontrivial 
junctions of vortices, 
like domain wall webs~\cite{Eto:2005cp} 
in subsection \ref{sec.junction} 
(or $(p,q)$ string webs). 
One negative circumstantial evidence is the fact 
that this system does not contain junction charge 
unlike domain wall junction with the Hitchin charge
or intersecting vortices with the instanton charge.
However this may not be crucial because 
wall junctions and intersecting vortices of the type I\!Ib
exist in the strong gauge coupling limit 
in which these junction charges vanish.

%%%%%%%%%%%%%%%%%%%%%%%%%%%%%%%%%%%%%%%%%%%%%%%%%%%%%%%%%%%%%%%%%%
\subsubsection{Vortex - Wall}
%%%%%%%%%%%%%%%%%%%%%%%%%%%%%%%%%%%%%%%%%%%%%%%%%%%%%%%%%%%%%%%%%%
\addtocontents{toc}{\vspace{-2.3mm}}

By performing the SS dimensional reduction along the $x^3$-direction
in the above 1/4 BPS system of three orthogonal vortices, we can find
another set of new 1/4 BPS equations for vortices and walls, given by
\begin{eqnarray}
&& F_{12}=\xi _3Y_3,\quad {\cal D}_1\Sigma _3= \xi _2Y_2,\quad 
{\cal D}_2\Sigma _3=-\xi _1Y_1,\label{vww1}\\
&& \sum_n (\tau_n)^i{_j}\D_n 
H^{jA}= i\tau_3
\left[(\Sigma _3)^A{}_BH^{jB}-m_{3,A}H^{jA}\right],\label{vww2}
\end{eqnarray}
with $n=1,2$.
The configuration can be summarized as follows
\begin{eqnarray*}
\begin{array}{c|ccccc|ccc}
{\rm VWW} &0 &1 &2 &(4) &(5) & & {\rm FI} &{\rm sign}\\  \hline
{\rm Wall}&\bigcirc &\bigcirc &\times &\bigcirc &\bigcirc &i\sigma _1& \zeta_1^I&\xi _1\\
{\rm Wall}&\bigcirc &\times &\bigcirc &\bigcirc &\bigcirc &i\sigma _2& \zeta_2^J&\xi _2\\
{\rm Vortex}&\bigcirc &\times &\times &\bigcirc &\bigcirc &i\sigma _3& \zeta_3^K&\xi _3
\end{array}
\end{eqnarray*} 
The energy of these composite states can be calculated as 
\begin{eqnarray}
E_{\rm m}=\xi _3{\cal V}_{12}^{(3)}
+ \left(\xi _1{\cal W}_{2,3}^{(1)}\right)v_{1}
+ \left(-\xi _2{\cal W}_{1,3}^{(2)}\right)v_{2}.
\end{eqnarray}

Vortex-strings in this system are parallel to both 
walls perpendicular to $x^1$ and those to $x^2$. 
This system is very similar to 
the system admitting wall junctions discussed in 
subsection \ref{sec.junction}. 
In the previous case, 
codimension two objects are the Hitchin vortices, 
which can live only in the unbroken phase. 
They are able to be localized at junction lines of walls,  
because gauge group is almost recovered inside walls.
They can play a role of binding energy of walls. 
However codimension two objects in the current case 
are ordinary vortices which live in the Higgs phase. 
Therefore they cannot come inside walls but 
rather live outside walls. 
Since they cannot become binding energy,  
we do not know if there can exist nontrivial wall junctions 
in this case; this system may admit only trivially 
(or at most asymptotically) orthogonally intersecting walls. 
However at least in the strong gauge coupling limit, 
the junction charge vanishes in the previous case. 
This implies that junction charge is not needed 
to make wall junctions in general. 
Therefore some nonlinear sigma models 
may admit wall junctions 
with vortex(lump)-strings outside walls.

%%%%%%%%%%%%%%%%%%%%%%%%%
\subsection{Summary of 1/4 BPS systems}
%%%%%%%%%%%%%%%%%%%%%%%%%%%%%%%%%%%%%%%%%%%%%%%%%%%%%%%%%%%%%%%%%%
\addtocontents{toc}{\vspace{-4mm}}

In this section we have shown that all 1/4 BPS states 
are classified into two classes, 
the types I\!Ia and I\!Ib, 
which preserve ${\cal N}=(1,1)$ and $(2,0)$, respectively, 
in the case of two dimensional world-volume.
According to this classification, 
there exist two sets of 1/4 BPS equations in $d=6$. 
We have obtained diverse 1/4 BPS equations 
in dimensions less than six. 
They admit various composite solitons 
as summarized in Table~\ref{1/4BPS-table}. 
Most of the 1/4 BPS equations in the type I\!Ib have been known already, 
whereas 
those in  the type I\!Ia have not been found yet and are new. 
\begin{table}[ht]
\begin{center}
\caption{1/4 BPS states. Definitions of all symbols here are given
in the caption of Table \ref{summary_table}. }
\label{1/4BPS-table}
{\small
\begin{tabular}{|c|cc|c|c|c|c|c|} \hline
&&& $d=6$&$d=5$&$d=4$&$d=3$&$d=2$\\
&&& ${\cal N}=1$ & ${\cal N}=1$ & ${\cal N}=2$&
   ${\cal N}=4$&${\cal N}=(4,4)$ \\ \hline\hline
type I\!Ib&$\tilde d=2,$&$ {\cal N}=(2,0)$& (w-)IVV&
   (w-)MVW&\begin{tabular}{c}      
(w-)HWW\\(w-)HV
\end{tabular}&(w-)NW &- \\
&$\tilde d=1$,&${\cal N}=2$ &
  -&(Q-)IVV&(Q-)MVW&
\begin{tabular}{c}      
(Q-)HWW\\(Q-)HV
\end{tabular}&(Q-)NW\\ \hline
&$\tilde d=3$,&${\cal N}=1$&  VVV & VWW & WW &-&-\\
type I\!Ia&$\tilde d=2$,&$ {\cal N}=(1,1)$
  &-&VVV&VWW& WW&-\\
&$\tilde d=1$,&${\cal N}=2$& -&-&VVV&VWW&WW\\
\hline
\end{tabular}
}
\end{center}
\end{table}

Like the type I\!Va and I\!Vb in 1/2 BPS states, 
the type I\!Ia and type I\!Ib exhibit very different features. 
Since composite states contain 
solitons with different dimensions, 
it depends if there exist normalizable zero modes or not. 
If they exist dynamics of these composite 
states can be described by nonlinear sigma models 
on the target space as the moduli space.  
According to the unbroken SUSY which they preserve, 
their target spaces as moduli spaces are Riemann manifolds 
or K\"ahler manifolds (with torsion) \cite{Hull:1985jv} 
for the type I\!Ia or I\!Ib, respectively. 

By the group including the present authors, 
the implicit solutions for the most of the 1/4 BPS equations of 
the type I\!Ib were obtained except for HV and NW in the table. 
For instance  
MVW was solved in \cite{INOS3}, 
IVV in \cite{Eto:2004rz} and 
HWW (wall webs) in \cite{Eto:2005cp,Eto:2005fm}.
All the exact solutions are available in the strong gauge coupling limit. 
For some cases, the moduli spaces were determined 
and explicit relations between moduli parameters and 
actual soliton configurations were examined. 
On the other hand, 
no solutions are found yet for 
the 1/4 BPS equations of the type I\!Ia.

%%%%%%%%%%%%%%%%%%%%%%%%%%%%%%%%%%%%%%%%%%%%%%%%%%%%%%%%%%%%%%%%%%%%%
\section{1/8 BPS Systems: the type I} \label{sect.5}
%%%%%%%%%%%%%%%%%%%%%%%%%%%%%%%%%%%%%%%%%%%%%%%%%%%%%%%%%%%%%%%%%%%%%
\addtocontents{toc}{\vspace{-2.3mm}}

The results in this section are new. 

%%%%%%%%%%%%%%%%%%%%%%%%%%%%%%%%%%%%%%%%%%%%%%%%%%%%%%%%%%%%%%%%%%%%%
\subsection{Instanton - Vortex}
%%%%%%%%%%%%%%%%%%%%%%%%%%%%%%%%%%%%%%%%%%%%%%%%%%%%%%%%%%%%%%%%%%%%%
\addtocontents{toc}{\vspace{-2.3mm}}

As we mentioned in Sec.\ref{sect.3}, the fundamental 1/2 BPS solitons
in the 6 dimensions with ${\cal N}=1$ SUSY 
are the vortex (3-brane) and the instanton (1-brane)
like M2 and M5-brane which are fundamental 1/2 BPS objects in M-theory.
In the previous section, 
we have found that two composite states of the vortices (V)
and the instantons (I), namely 
the type I\!Ia (VVV) and the type I\!Ib (IVV), 
are possible as the 1/4 BPS states. 
Furthermore, there exists the unique 1/8 BPS composite state
of the vortices and the instantons (the type I).
A set of the projection operators
on the Killing spinor, 
$\Gamma_a\varepsilon = \lambda_a\varepsilon$, 
is given as follows
\begin{eqnarray}
\Gamma _a  = \left\{ \gamma ^{05},\,
\gamma ^{12}i\sigma _3,\,\gamma ^{34}i\sigma _3,\,
\gamma ^{23}i\sigma _1,\,\gamma ^{14}i\sigma _1,\,
\gamma ^{31}i\sigma _2,\,\gamma ^{24}i\sigma _2
 \right\},
\label{proj_1/8}
\end{eqnarray}
where we assign signs
$
\lambda _a = \left(\eta ,\,\xi _3,\,-\eta \xi _3,\,\xi _1,\,-\eta \xi _1,\,
\xi _2,\,-\eta \xi _2\right)$
with $\xi _1\xi _2\xi _3=-1$.
Requiring the invariance $\delta \lambda^i = 0$ and 
$\delta \psi^A = 0$ in Eq.(\ref{susy_transf}) for the 1/8 SUSY
projected by the above operators, we get the 
following 1/8 BPS equations
\begin{eqnarray}
 && \myfrac12\epsilon _{nml}F_{ml}-\eta F_{n4} 
    = \xi _nY_n,\quad (m,n,l = 1,2,3), \label{typeIeq0}\\
 && \sum_{m=1}^4(q_m)^i{_j}\D_mH^j=0 , \label{typeIeq} \\
 && q_m \equiv (\xi_1i\sigma_1,\xi_2i\sigma_2,\xi_3i\sigma_3,-\eta{\bf 1}_2) .
\end{eqnarray}
It may be interesting to note the quaternionic structure $q_m$ in 
these equations.
% where $q_m = (\xi_1i\sigma_1,\xi_2i\sigma_2,\xi_3i\sigma_3,-\eta{\bf 1}_2)$.
We summarize this 1/8 BPS configuration below.
\begin{eqnarray*}
\begin{array}{c|cccccc|cc}
{\rm IV^6}&0 &1 &2 &3 &4 &(5) &  &{\rm sign}\\ \hline
{\rm Instanton}&\bigcirc &\times &\times &\times &\times &\bigcirc &{\bf 1}_2&\eta \\
{\rm Vortex}&\bigcirc &\bigcirc &\times &\times &\bigcirc &\bigcirc &i\sigma _1 &\xi _1\\
{\rm Vortex}&\bigcirc &\times &\bigcirc &\times &\bigcirc &\bigcirc &i\sigma _2 &\xi _2\\
{\rm Vortex}&\bigcirc &\times &\times &\bigcirc &\bigcirc &\bigcirc &i\sigma _3 &\xi _3\\
{\rm Vortex}&\bigcirc &\times &\bigcirc &\bigcirc &\times &\bigcirc &i\sigma _1&-\eta \xi _1\\
{\rm Vortex}&\bigcirc &\bigcirc &\times &\bigcirc &\times &\bigcirc &i\sigma _2&-\eta \xi _2\\
{\rm Vortex}&\bigcirc &\bigcirc &\bigcirc &\times &\times &\bigcirc &i\sigma _3&-\eta \xi _3
\end{array}
\end{eqnarray*} 
This 1/8 BPS configurations of the instantons and the vortices
are independent on $x^5$, so these also exist in 5 dimensions.
This configuration without instantons was 
expected in \cite{Portugues:2002ih} without any equations 
in the context of hyper-K\"ahler nonlinear sigma models in $d=5$.

Since the set of the projection operators in Eq.(\ref{proj_1/8})
contains $\gamma^{05}$, we can add Eq.(\ref{wave_SD})
to the above 1/8 BPS equations still preserving the same 1/8 SUSY.
Similarly to the type I\!Vb and the type I\!Ib, Eq.(\ref{wave_SD}) gives 
wavy extensions, and the dyonic extension when we perform the 
trivial dimensional reduction along the $x^5$-direction.
Then the energy of this composite state is 
the sum of energies of instantons and vortices 
with energy coming from the electric flux
\begin{eqnarray}
E&=&-\eta X+\eta {\cal I}_{1234}
+ \left( \xi_3 {\cal V}_{12}^{(3)}\right)v_{34}
+ \left( \xi_1 {\cal V}_{23}^{(1)}\right)v_{14}
+ \left( \xi_2 {\cal V}_{31}^{(2)}\right)v_{24}\nonumber\\
&&- \left( \eta \xi_3 {\cal V}_{34}^{(3)}\right)v_{12}
- \left( \eta \xi_1 {\cal V}_{14}^{(1)}\right)v_{23}
- \left( \eta \xi_2 {\cal V}_{24}^{(2)}\right)v_{13}, 
\end{eqnarray}
where $X$ denotes the 5th component of the Poynting vector 
(\ref{S5'})
or the electric charge (\ref{S5''}) of the dyonic solitons.

%%%%%%%%%%%%%%%%%%%%%%%%%%%%%%%%%%%%%%%%%%%%%%%%%%%%%%%%%%%%%%%%%%
\subsection{Descendants of the type I}
%%%%%%%%%%%%%%%%%%%%%%%%%%%%%%%%%%%%%%%%%%%%%%%%%%%%%%%%%%%%%%%%%%

%%%%%%%%%%%%%%%%%%%%%%%%%%%%%%%%%%%%%%%%%%%%%%%%%%%%%%%%%%%%%%%%%%
\subsubsection{Monopole - Vortex - Wall}
%%%%%%%%%%%%%%%%%%%%%%%%%%%%%%%%%%%%%%%%%%%%%%%%%%%%%%%%%%%%%%%%%%

By performing the SS dimensional reduction along the $x^4$-direction,
the type I 1/8 BPS states in 6 dimensions given in the previous
subsection reduce to other  1/8 BPS composite states which
contain monopoles, vortices and walls.
The corresponding 1/8 BPS equations are of the form
\begin{eqnarray}
&&\myfrac12\epsilon _{nml}F_{ml}+\eta {\cal D}_n\Sigma _4=\xi _nY_n,
 \label{typeI'}\\
&&\sum_{n=1}^3(\tau_n)^i{}_j{\cal D}_nH^j=
-i\eta \left[(\Sigma _4)^A{}_BH^{iB}-m_{4,A}H^{iA}\right] .
 \label{typeI''}
\end{eqnarray}
The configurations are summarized in the following table.
\begin{eqnarray*}
\begin{array}{c|ccccc|cc}
{\rm MV^3W^3}&0 &1 &2 &3 &(5) &  &{\rm sign}\\  \hline
{\rm Monopole}&\bigcirc &\times &\times &\times &\bigcirc &{\bf 1}_2&\eta \\
{\rm Vortex}&\bigcirc &\bigcirc &\times &\times &\bigcirc &i\sigma _1 &\xi _1\\
{\rm Vortex}&\bigcirc &\times &\bigcirc &\times &\bigcirc &i\sigma _2 &\xi _2\\
{\rm Vortex}&\bigcirc &\times &\times &\bigcirc &\bigcirc &i\sigma _3 &\xi _3\\
{\rm Wall}&\bigcirc &\times &\bigcirc &\bigcirc &\bigcirc &i\sigma _1&-\eta \xi _1\\
{\rm Wall}&\bigcirc &\bigcirc &\times &\bigcirc &\bigcirc &i\sigma _2&-\eta \xi _2\\
{\rm Wall}&\bigcirc &\bigcirc &\bigcirc &\times &\bigcirc &i\sigma _3&-\eta \xi _3
\end{array}
\end{eqnarray*}  
Their energy is given by the sum of contributions from 
monopoles, vortices and walls, 
and energy coming from the electric flux or the electric charge: 
\begin{eqnarray}
 E&=&-\eta X + \eta {\cal M}_{123}
+ \left( \xi_3 {\cal V}_{12}^{(3)}\right)v_3
+ \left( \xi_1 {\cal V}_{23}^{(1)}\right)v_1
+ \left( \xi_2 {\cal V}_{31}^{(2)}\right)v_2\nonumber\\
&&- \left( \eta \xi_3 {\cal W}_{3,4}^{(3)}\right)v_{12}
- \left( \eta \xi_1 {\cal W}_{1,4}^{(1)}\right)v_{23}
- \left( \eta \xi_2 {\cal W}_{2,4}^{(2)}\right)v_{31}
\end{eqnarray}
with $X$ denoting the 5th component of the Poynting vector 
(\ref{S5'})
or the electric charge (\ref{S5''}) of the dyonic solitons.

At this stage we cannot expect what kinds of configuration  are actually
possible as solutions of the 1/8 BPS equations (\ref{typeI'})
and (\ref{typeI''}). 
As one of interesting possibilities, 
we expect that there may exist 
a 1/8 BPS state of a vortex-string network.
Such expectation seems to be natural 
because it has been well established that 
two vortices extending to opposite directions 
can be connected at a monopole junction point, 
as a 1/4 BPS state of a monopole in the Higgs phase~\cite{vm}.
Since this 1/8 BPS system contains 
monopoles and vortices extending to different three directions, 
two or more vortices may meet at a monopole junction
point, and junctions may constitute 
a three-dimensional vortex-string network, 
not like two-dimensional $(p,q)$ string webs.

%%%%%%%%%%%%%%%

%%%%%%%%%%%%%%%%%%%%%%%%%%%%%%%%%%%%%%%%%%%%%%%%%%%%%%%%%%%%%%%%%%
\subsubsection{Hitchin vortex - Vortex - Wall}
%%%%%%%%%%%%%%%%%%%%%%%%%%%%%%%%%%%%%%%%%%%%%%%%%%%%%%%%%%%%%%%%%%

When we further dimensionally reduce the $x^3$-direction, 
we obtain a system of 
vortices and Hitchin vortices both living in the $x^1$-$x^2$ plane 
and walls perpendicular to 
the $x^2$ direction 
and walls to the $x^1$ direction  
in $d=4$, ${\cal N}=2$ gauge theory. 
The BPS equations for this system can be obtained as 
the dimensional reduction of 
the BPS equations (\ref{typeI'}) and (\ref{typeI''}), 
to give 
\begin{eqnarray}
\D_1\Sigma_3 + \eta \D_2\Sigma_4 = \xi_2Y_2,\quad
- \D_2\Sigma_3 + \eta \D_1\Sigma_4 = \xi_1 Y_1,\quad
F_{12} - i\eta\left[\Sigma_3,\Sigma_4\right]
= \xi_3 Y_3,
\label{hvww1}\\
\sum_{n=1}^2 (\tau_n)^i{_j}\D_nH^j
= -\xi_3(\sigma_3)^i{_j}\left[(\Sigma_3)^A{_B}H^{jB} - m_{3,A}H^{jA}\right]
-i\eta \left[ (\Sigma_4)^A{_B}H^{iB} - m_{4,A}H^{iA}\right].
\label{hvww2}
\end{eqnarray}
The energy of these 1/8 BPS states is given by
\begin{eqnarray}
 E&=&-\eta X + \eta {\cal H}_{12} + \xi_3 {\cal V}_{12}^{(3)}\nonumber\\
&&+ \left( \xi_1 {\cal W}_{2,3}^{(1)} - \eta \xi_2 {\cal W}_{2,4}^{(2)}\right)v_1
- \left( \xi_2 {\cal W}_{1,3}^{(2)} 
+ \eta \xi_1 {\cal W}_{1,4}^{(1)}\right)v_2
\end{eqnarray} 
with $X$ denoting the 5th component of the Poynting vector 
(\ref{S5'})
or the electric charge (\ref{S5''}) of the dyonic solitons.
The configurations are summarized in the following table.
\begin{eqnarray*}
\begin{array}{c|cccc|cc}
{\rm HVW^2W^2}&0 &1 &2 &(5) &  &{\rm sign}\\  \hline
{\rm Hitchin\; vortex}&\bigcirc &\times &\times  &\bigcirc &{\bf 1}_2&\eta \\
{\rm Wall}&\bigcirc &\bigcirc &\times  &\bigcirc &i\sigma _1 &\xi _1\\
{\rm Wall}&\bigcirc &\times &\bigcirc  &\bigcirc &i\sigma _2 &\xi _2\\
{\rm Vortex}&\bigcirc &\times &\times  &\bigcirc &i\sigma _3 &\xi _3\\
{\rm Wall}&\bigcirc &\times &\bigcirc  &\bigcirc &i\sigma _1&-\eta \xi _1\\
{\rm Wall}&\bigcirc &\bigcirc &\times  &\bigcirc &i\sigma _2&-\eta \xi _2
\end{array}
\end{eqnarray*}  

The 1/8 BPS equations (\ref{hvww1}) and (\ref{hvww2})
were eventually obtained in \cite{Kakimoto:2003zu} 
in the case of $U(1)$ gauge theory 
coupled to two hypermultiplets with equal $U(1)$ charges, 
although the authors in \cite{Kakimoto:2003zu} did not 
realize that they are 1/8 BPS 
and just analyzed 1/4 BPS wall junction 
(with Hitchin vortices) without ordinary vortices  
at that time. 
More general solutions would become 
wall junctions with vortices outside walls, 
but it is difficult to construct them at this stage.  

Another interesting possibility of solitons 
is the Hitchin vortex inside vortices, 
both extending to fifth direction.   
For concreteness let us consider 
$U(2)$ gauge theory with two hypermultiplets 
in the fundamental representation 
with equal $U(1)$ charges.  
Since the Hitchin vortex live in 
unbroken phase (of non-Abelian gauge group), 
it cannot live inside a single vortex, 
where only $U(1)$ gauge group is recovered.
If positions of two vortices eventually 
coincide but their orientations are in 
the upper left 
and the lower right elements, 
$U(2)$ gauge group is recovered 
at that coincident point (line). 
Then we expect that a Hitchin vortex 
lives inside those two coincident vortices 
with gluing them together, 
to become a 1/8 BPS composite state of 
the Hitchin vortex and two vortices. 

\medskip
When we perform the further SS dimensional reduction 
along the $x^2$-direction, we get another set of 
1/8 BPS equations in $d=3$ $(2)$ 
which contains the Nahm walls and walls (NW$^3$).

Before closing this section let us give a comment about 
supersymmetry in $d=4$. After the SS dimensional reductions
twice along the $x^4$ and $x^3$-directions from $d=6$, the projection operator
$\gamma^{34}i\sigma_3$ does not give any restriction for 
the BPS equations and just separate eight supercharges to 2 sets of 
four supercharges.
This means that the 1/8 BPS equations (\ref{hvww1}) and (\ref{hvww2})
can be embedded into the model with ${\cal N}=1$ in 4 dimensions
as a 1/4 BPS equations.

%%%%%%%%%%%%%%%%%%%%%%%%%%%%%%%%%%%%%
\subsection{Summary of $1/8$ BPS systems}
\addtocontents{toc}{\vspace{-4mm}}

Since 1/8 BPS states preserve just one SUSY,  
a set of 1/8 BPS equations is the unique in the maximal dimension $d=6$. 
All 1/8 BPS equations in dimension less than six 
can be obtained from it by dimensional reductions. 
We have written down BPS equations in $d=5$ and $d=4$, 
but we have stopped it there because it is a simple task. 
All 1/8 BPS system in gauge theory with eight supercharges 
are summarized in Table \ref{1/8BPS-table}. 
One can find the rests of 1/8 BPS systems in lower dimensions. 

\begin{table}[ht]
\begin{center}
\caption{1/8 BPS states. Definitions of all symbols here are given
in the caption of Table \ref{summary_table}. }
\label{1/8BPS-table} 
{\small
\begin{tabular}{|c|c|c|c|c|c|} \hline
& $d=6$& $d=5$& $d=4$& $d=3$& $d=2$\\
& ${\cal N}=1$ & ${\cal N}=1$ & ${\cal N}=2$&
${\cal N}=4$&${\cal N}=(4,4)$ \\ \hline\hline
$\tilde d=2$&(w-)IV$^6$&(w-)MV$^3$W$^3$
&(w-)HVW$^2$W$^2$&(w-)NW$^3$& -\\
$\tilde d=1$&-&(Q-)IV$^6$&(Q-)MV$^3$W$^3$
&(Q-)HVW$^2$W$^2$&(Q-)NW$^3$\\
\hline
\end{tabular}
}
\end{center}
\end{table}

Since all 1/2 or 1/4 BPS equations can be obtained 
by simply ignoring some projections, 
their solutions automatically satisfy 
the unique 1/8 BPS equations (\ref{typeIeq0}) and (\ref{typeIeq}).
Therefore 1/8 BPS equations contain 
all possible BPS states in theories with eight supercharges. 
However, unlike the cases of 1/2 or 1/4 BPS states, 
solving 1/8 BPS equations is very difficult. 
It is still difficult to find even a special 1/8 BPS solution. 
That remains as a future problem.

%%%%%%%%%%%%%%%%%%%%%%%%%%%%%%%%%%%

\section{Conclusion and Discussion}

We have systematically derived 
1/4 BPS and 1/8 BPS equations 
describing composite states made of various 
1/2 BPS solitons in 
SUSY gauge theories in $d=6,5,4,3,2$ with eight supercharges.
First of all we have derived, in 6 dimensions, 
all the 1/2 BPS, 1/4 BPS and 
1/8 BPS equations by clarifying 
the projection operators giving the Killing conditions 
on fermions of unbroken supersymmetry transformation.  
Then we have found their descendant BPS equations in lower dimensions 
by performing the SS and/or 
the trivial dimensional reductions. 
While we have found lots of new BPS equations, 
we have also rederived known BPS equations. 
Let us briefly summarize which are new contribution 
found in this paper and which are not. 

We have found that 
the known 1/2 BPS objects 
in SUSY gauge theories with eight supercharges 
can be classified into two classes, 
according to the chirality of unbroken supercharges: 
the type I\!Va defined by the projections (\ref{4a}) 
and the type I\!Vb defined by the projections (\ref{4b}), 
which preserve 
${\cal N}=(2,2)$ (non-chiral) SUSY 
and ${\cal N}=(4,0)$ (chiral) SUSY, respectively,  
in the case of two dimensional world-volume. 
The type I\!Va contains 
the vortex equations (\ref{eq:bps_vor1}) and (\ref{eq:bps_vor2}) 
studied in Refs.~\cite{ANO}--\cite{Eto:2006uw}
and domain-wall equations (\ref{eq:bps_wal1}) and (\ref{eq:bps_wal2})
studied in Refs.~\cite{Abraham:1992vb}--\cite{Hanany:2005bq}.
The type IVb contains the well established soliton equations: 
the instanton equations (\ref{eq:bps_inst}), 
the monopole equations (\ref{eq:monopole}), 
the Hitchin equation (\ref{eq:bps_hitc}) 
and the Nahm equation (\ref{eq:bps_nahm}).
As a new result, we have found an extension of solitons in the type IVb 
to wavy solitons (\ref{eq_05}), along the $x^5$-direction of whose 
world-volume electric waves travel. 
We have constructed concrete solutions (\ref{eq:sol_wavy}) of 
these wavy solitons.
They are reduced to 
known dyonic solitons (\ref{eq:bps_dyon}) 
by the dimensional reduction along the $x^5$-direction.
The dyonic instantons \cite{Lambert:1999ua} and dyons \cite{Julia:1975ff}
are such solitons.

As to 1/4 BPS objects, we have classified them into 
the type I\!Ia defined by the projections (\ref{type A}) and 
the type IIb defined by the projections (\ref{type B}) 
which preserve ${\cal N}=(1,1)$ (non-chiral) SUSY 
and ${\cal N}=(2,0)$ (chiral) SUSY, respectively,  
in the case of two dimensional world-volume.
Recently much attention have been paid to the type I\!Ib solitons: 
the vortex-instanton system described by Eqs.(\ref{ivv1}) and (\ref{ivv2})
was analyzed in \cite{Eto:2004rz}, 
the monopole-vortex-domain-wall system described by 
Eqs.(\ref{wvm1}) and (\ref{wvm2}) 
was analyzed
in Refs.~\cite{Gauntlett:2000de}--\cite{Sakai:2005sp}
and the system of domain wall junctions (\ref{jww1})-(\ref{jww4}) 
was analyzed in
Refs.~\cite{Kakimoto:2003zu}--\cite{Eto:2005mx}.
We also have discussed in Sec.~\ref{HVNW} 
possibility of 
new composite states made of Hitchin vortices and ordinary vortices, 
and of Nahm walls and domain walls. 
Besides the descent relations between these 1/4 BPS objects, 
we have pointed out that topological charges in the type I\!Vb, 
namely the instanton charge, the monopole charge and the Hitchin charge, 
arise at junction points of different solitons. 
In the Abelian gauge theory the junction charges contribute 
negative masses to the total energy 
while the topological charges with positive masses arise 
in the non-Abelian gauge theory as usual. 
As a new result, we have found that the type I\!Ib 
has the wavy and the dyonic extensions (\ref{wavy_1/4}) 
with keeping the 1/4 BPS condition, 
much like solitons of the type I\!Vb. 

On the other hand, unlike the type I\!Ib, 
the type I\!Ia includes 
many new interesting equations which have not been studied yet. 
The new equations (\ref{vvv1}) and (\ref{vvv2}) describe 
intersecting vortex-strings extending to three different directions.
These equations reduce in the strong gauge coupling limit 
to those for intersecting lump-strings in 
a hyper-K\"ahler sigma model \cite{Naganuma:2001pu,Portugues:2002ih}. 
The other new equations (\ref{vww1}) and (\ref{vww2}) 
describe composite states of 
intersecting domain walls and 
vortex-strings which sit outside walls and 
are parallel to the intersection line. 
For both sets of equations, constructing solutions is not succeeded yet and 
remains as a future problem. 

Finally, we have explored 
the unique set of the 1/8 BPS equations (\ref{typeIeq0}) and 
(\ref{typeIeq}) in six dimensions from 
the type I projections (\ref{type-I}).  
These equation describe 
instanton-strings and vortex-3-branes 
in six different directions. 
These equations (\ref{typeIeq0}) and (\ref{typeIeq}) are completely new and 
are nicely written in terms of the quaternion.  
We then have obtained the 1/8 BPS descendants of 
Eqs.~(\ref{typeIeq0}) and (\ref{typeIeq}): 
the equations (\ref{typeI'}) and (\ref{typeI''}) describing 
monopoles with triply intersecting vortices and 
triply intersecting domain walls, 
and the equations (\ref{hvww1}) and (\ref{hvww2}) 
describing two kinds of intersecting domain walls with 
the Hitchin charge and vortices outside walls.  
We have pointed out a three-dimensional 
network of vortex-strings connected by monopoles 
as a possible solution of 
Eqs.~(\ref{typeI'}) and (\ref{typeI''}). 
We have shown that 
all of these 1/8 BPS equations have the wavy and the dyonic extensions. 
We can call the equations (\ref{typeIeq0}) and (\ref{typeIeq}) 
as the mother equations of 
the Yang-Mills-Higgs system in the sense 
that all the other BPS equations including 1/2 BPS and 1/4 BPS equations 
can be derived from these equations by the SS and/or
the trivial dimensional reductions and/or  
by truncating appropriate coupling constants and/or fields. 
Solving the mother equations (\ref{typeIeq0}) and (\ref{typeIeq}) 
or determining their moduli space is one of final goal of this subject.

\medskip
We present several discussion and future directions here.

{\it D-brane configurations.} 
In this paper we have considered general gauge group 
with general matter contents. 
The mostly discussed model 
is the $U(\NC)$ gauge theory 
coupled to hypermultiplets 
in the fundamental representation 
with equal $U(1)$ charge.  
This model can be realized 
as the effective gauge theory on 
$\NC$ D$p$-branes under the background 
of $\NF$ D($p+4$)-branes separated to 
one, two, three and four directions, 
corresponding to real, complex, triplet and quartet masses 
for $p=1,2,3,4$, respectively.  
In that model, 
1/2 BPS walls are realized 
as $\NC$ kinky D$p$-branes traveling 
between $\NF$ separated  
D($p+4$)-branes~\cite{Lambert:1999ix,Eto:2004vy}.  
In the case of 1/4 BPS wall junctions \cite{Eto:2005cp} 
in Sect.\ref{HVNW}, 
two of transversal directions of D3-branes, 
in a plane orthogonal to D7-branes with complex positions, 
depend on two world-volume directions D3-branes \cite{Eto:2005mx}. 
Brane configurations for vortices 
were given by Hanany and Tong~\cite{vortices}. 
Duality between vortices and domain walls is understood as 
T-duality between corresponding brane configurations \cite{Eto:2006mz}.
Brane configuration for a monopole in 
the Higgs was also obtained by Hanany and Tong~\cite{vm}.
Diverse BPS equations found in this paper 
will provide more variations 
of ample D-brane configurations.

{\it The Nahm transformation between 
solutions of vortices and walls and between 
composite states.} 
The type I\!Vb equations are the members of 
the well-established series of 
instantons (SDYM), BPS monopoles, 
the Hitchin vortices, 
the Nahm walls (the Nahm data), 
if we take the Kaluza-Klein modes into account. 
The Nahm transformation maps between 
the ADHM/Nahm data and instanton/monopole solutions.  
Therefore we expect 
that there should exist 
a Nahm-like transformation 
among the type I\!Va equations 
for vortices and domain walls.
We also suspect similar relation 
for the type I\!Ib equations describing, 
for instance, instantons/monopoles in the Higgs phase.

{\it Higher co-dimensions/dimensions and higher SUSY}. 
The same analysis should be applied 
to theories with sixteen supercharges. 
The maximal dimension is ten in this case. 
Therefore BPS solitons with higher codimensions 
and/or in higher dimensions are possible. 
For instance a BPS soliton of co-dimension five 
was found in \cite{Kihara:2004yz}. 
The Donaldson-Uhlenbeck-Yau instantons \cite{DUY}
are of co-dimension six and are 
1/4 BPS in $d=7$~\cite{Bak:2002aq,Townsend:2004nc}.\footnote{
It has been recently pointed out in \cite{Popov:2005ik} 
that the SDYM-Higgs equations (\ref{ivv1}) and (\ref{ivv2}) 
in $d=4+1$ can be derived from 
the Donaldson-Uhlenbeck-Yau equations in $d=6+1$  
by the $SU(2)$ equivariant dimensional reduction 
on $S^2$, at least in the case of $U(1)$ gauge group. 
} 
The octernionic instantons are of 
co-dimension eight~\cite{Corrigan:1982th, Grossman:1984pi}. 
Classification of these BPS solitons of higher codimensions 
is interesting extension.

{\it Non-commutative solitons}.
We have not considered solitons in non-commutative space. 
Extensions of BPS equations to 
non-commutative solitons should be possible.

%%%%%%%%%%%%%%%%%%%%%%%%%%%%%%%%%%%%%%%%
\section*{Acknowledgements}

We are pleased to express 
much appreciation to Norisuke Sakai for 
fruitful discussions and continuous encouragements.
We would like to thank Kimyeong Lee and Ho-Ung Yee 
for kindly informing us their results. 
We are also grateful to Kazutoshi Ohta and Nobuyoshi Ohta 
for useful comments.  
The work of K.~O. and M.~N. (M.~E. and Y.~I.) is
supported by Japan Society for the Promotion 
of Science under the Post-doctoral (Pre-doctoral) Research Program.  
M.~N. wishes to thank KIAS for their hospitality.

%%%%%%%%%%%%%%%%%%%%%%%%%%%%%%%%%%%%%%%%%%%%%%%%%%%%%%%%%%%%%%%%%%%%%%%%%%%%%%%%

%%%%%%%%%%%%%%%%%%%%%%%%%%%%%%%%%%%%%%%%%%%%%%%%%%%%%%%%%%%%%%%%%%%

\end{document}